\documentclass[twocolumn,english,aps,superscriptaddress,prx]{revtex4-2}
\usepackage[utf8]{inputenc}

\usepackage{graphics}
\usepackage{color}
\usepackage{graphicx}
\usepackage{amssymb}

\usepackage{bm}

\usepackage{amsfonts,hyperref}
\usepackage{amsmath}
\usepackage{gensymb}

\newenvironment{Author Contributions}{%
   \section*{Author Contributions}%
   \setlength{\parskip}{0pt}%
   \small%
    }{}

\newcommand{\scbo}{SrCu$_{2}$(BO$_{3}$)$_{2}$}
\newcommand{\mgscbo}{SrCu$_{2-x}$Mg$_{x}$(BO$_{3}$)$_{2}$}

\renewcommand{\figurename}{{\bf{Figure}}}

\makeatletter
\makeatletter \renewcommand{\fnum@figure}{{\bf{\figurename~\thefigure}}}
\makeatother

\begin{document}

\title{Phase diagram of the Shastry-Sutherland Compound \scbo ~under extreme combined conditions of field and pressure}

\author{Zhenzhong Shi}
\affiliation{Department of Physics, Duke University, Durham, North Carolina 27708, USA}
\author{Sachith Dissanayake}
\affiliation{Department of Physics, Duke University, Durham, North Carolina 27708, USA}
\author{Philippe Corboz}
\affiliation{Institute for Theoretical Physics and Delta Institute for Theoretical Physics, University of Amsterdam, Science Park 904, 1098 XH Amsterdam, The Netherlands}
\author{William Steinhardt}
\affiliation{Department of Physics, Duke University, Durham, North Carolina 27708, USA}
\author{David Graf}
\affiliation{National High Magnetic Field Laboratory, Florida State University,Tallahassee, Florida 32310, USA}
\author{D.M. Silevitch}
\affiliation{Division of Physics, Math, and Astronomy, California Institute of Technology,Pasadena, California 91125, USA}
\author{Hanna A. Dabkowska}
\affiliation{Brockhouse Institute for Material Research, McMaster University, Hamilton, Ontario L8S 4M1, Canada}
\author{T.F. Rosenbaum}
\affiliation{Division of Physics, Math, and Astronomy, California Institute of Technology,Pasadena, California 91125, USA}
\author{Fr\'{e}d\'{e}ric Mila}
\affiliation{Institute of Physics, Ecole Polytechnique F\'{e}d\'{e}rale de Lausanne (EPFL), CH-1015 Lausanne, Switzerland}
\author{Sara Haravifard}
\email{sara.haravifard@duke.edu}
\affiliation{Department of Physics, Duke University, Durham, North Carolina 27708, USA}
\affiliation{Department of Mechanical Engineering and Materials Science, Duke University, Durham, North Carolina 27708, USA}
\date{\today}

\begin{abstract}
Motivated by the intriguing properties of the Shastry-Sutherland compound \scbo ~under pressure, with a still debated intermediate plaquette phase appearing at around 20 kbar and a possible deconfined critical point at higher pressure upon entering the antiferromagnetic phase, we have investigated its high-field properties in this pressure range using tunnel diode oscillator (TDO) measurements. The two main new phases revealed by these measurements are fully consistent with those identified by infinite Projected Entangled Pair states (iPEPS) calculations of the Shastry-Sutherland model, a 1/5 plateau and a $10 \times 2$ supersolid. Remarkably, these phases are descendants of the full-plaquette phase, the prominent candidate for the intermediate phase of \scbo. The emerging picture for \scbo ~is shown to be that of a system dominated by a tendency to an orthorhombic distortion at intermediate pressure, an important constraint on any realistic description of the transition into the antiferromagnetic phase.
\end{abstract}

\maketitle

\section{Introduction}

While the behavior of individual spins in isolation is well understood, complex behavior and new quantum states often emerge from networks of such spins, especially when competing interactions forestall the formation of simple ordered states, a phenomenon known as magnetic frustration~\cite{lacroix2011introduction}.  A key tool for understanding these states is the ability to tune parameters such as the relative strength of the different interactions or the external magnetic field. In that respect, the Shastry-Sutherland (SS) model, a 2-dimensional (2D) network of interacting spin dimers, together with its experimental realization \scbo, is a prominent candidate. Starting with the discovery of the first magnetization plateaus ~\cite{Kageyama1999} at 1/8 and 1/4 in 1999, and the confirmation that the translational symmetry is broken in the 1/8 plateau ~\cite{Kodama2002}, the interest in this layered frustrated quantum magnet and the related SS model has never decreased, and many new remarkable properties have been discovered since then. At ambient pressure, additional plateaus have been identified ~\cite{Onizuka2000,takigawa04,levy08,Sebastian2008,Jaime2012,Takigawa2013,matsuda13,Haravifard16} to build the improbable sequence 1/8, 2/15, 1/6, 1/4, 1/3, 2/5 and 1/2, and it took 15 years and the invention of tensor network algorithms to come up with a theory able to account for this remarkable series~\cite{corboz14_shastry}. Between the plateaus, translation invariance is never restored, and it remains a challenge to establish which of these intermediate phases are spin-supersolids and which are incommensurate phases with proliferating domain walls~\cite{Takigawa2013}. The excitation spectrum is also remarkable, with very flat bands, and it has been shown that, due to Dzyaloshinskii-Moriya interactions~\cite{Cepas01,Miyahara03,Kodama2005,Haravifard2014,romhanyi11}, a small field induces topological magnon bands with non-zero Chern numbers~\cite{Romhanyi2015} and experimental consequences still to be explored. Finally, this compound is remarkably sensitive to pressure for an oxide, and two phase transitions have been observed~\cite{Waki2007,Haravifard2012,Zayed2017,Sakurai2018,Guo2020,Boos2019a,Badrtdinov2020}. The first one is clearly first order, and it has been shown very recently that, as a function of temperature, it terminates at a critical point analogous to that of water~\cite{Jimenez2021}. 

Most of these properties are direct consequences of the peculiar arrangement of the Cu spins 1/2 in \scbo, which form weakly coupled 2D networks of orthogonal dimers topologically equivalent to SS model introduced in 1981~\cite{Shastry81}. For the pure Heisenberg model, the exact ground state is a product of singlet dimers~\cite{Shastry81,Miyahara99} as long as the inter-dimer coupling $J'$ is not too large as compared to the intra-dimer coupling $J$. Heuristically, we can think of the magnetization as due to magnetic particles $T_1$ that form when a dimer singlet $S$ is replaced by a triplet polarized along the field. These particles have a very small kinetic energy due to the highly frustrated dimer arrangement, leading to very flat bands and to Mott insulating phases (i.e. magnetization plateaus) at fractional fillings. Some of these plateaus (1/4, 1/3, 1/2) can be simply interpreted as Wigner crystals of $T_1$ particles, while the lower magnetization plateaus are best seen as Wigner crystals of spin-2 bound states that form because of a second-order kinetic term in $J'/J$ that leads to a binding between pairs of $T_1$ particles on neighboring parallel dimers~\cite{corboz14_shastry}. Additionally, the supersolid phases correspond to adding $T_1$ particles to a plateau phase, the hopping of these extra-particles being due to correlated hopping that takes advantage of the underlying network of $T_1$ particles~\cite{Momoi00,Schmidt2008}.

The properties under pressure are the consequence of another remarkable property of \scbo. The intra-dimer Cu-Cu bond is close to ninety degrees, and applying pressure brings this angle even closer to ninety degrees, reducing $J$ and increasing the ratio $J'/J$~\cite{Radtke2015,Badrtdinov2020}.  Now, the phase diagram of the SS model has three phases~\cite{Koga2000,Takushima01,Chung01,Laeuchli2002,Corboz13_shastry}: an exact dimer phase up to $J'/J\simeq 0.675$, an antiferromagnetic phase above $J'/J= 0.765(15)$~\cite{Corboz13_shastry} (in the limit $J^\prime/J  \longrightarrow  \infty $ the SS model is equivalent to the square lattice antiferromagnet), and an intermediate plaquette phase in between where strong $J'$ bonds form around half the empty square plaquettes of the SS lattice. The transition between the dimer phase and the intermediate phase is clearly first order, but the nature of the transition between the intermediate phase and the AF phase is still debated, and the interest in this transition has risen recently after the proposal that it could be a deconfined quantum critical point~\cite{Senthil2004,Lee2019a,Yang2021}. 

In \scbo, the ratio $J'/J\simeq 0.63$~\cite{matsuda13} puts it close to the boundary of the dimer phase, and indeed it turns out that a pressure of about 20 kbar is sufficient to induce a first-order transition into another gapped phase~\cite{Waki2007,Haravifard2012,Zayed2017,Sakurai2018,Guo2020}, followed at higher pressure by a transition into another phase still to be fully characterized~\cite{Guo2020,Jimenez2021}. Given the topology of the system, it is natural to expect the gapped phase to be the intermediate phase of the SS model, but this does not seem to be the case. NMR experiments have revealed early on that there are two types of Cu sites~\cite{Waki2007}, inconsistent with the intermediate phase of the SS model, and various experimental results seem to be rather consistent with a full plaquette phase where strong $J'$ bonds form around half the square plaquettes that contain a dimer, likely accompanied by an orthorhombic lattice distortion~\cite{Waki2007,Zayed2017,Boos2019a}. In the absence of direct probes of the symmetry of this intermediate phase, its precise nature remains an open issue, and a very important one because the nature of the intermediate phase will of course influence the nature of the transition into the AF phase. 

In this paper, our aim is to gain insight into the properties of \scbo ~under pressure by an investigation of its high-field properties in the relevant pressure range using Tunnel Diode Oscillator (TDO) measurements, and by an investigation of the high-field properties of the SS model in the corresponding $J'/J$ range using tensor network methods. These regions of the phase diagrams of \scbo ~and of the SS model have not been previously explored, and, as we demonstrate, the new phases that appear in high field shed new light on the problem and set the ground for further studies of \scbo ~under pressure. 

\section{Experimental Results}
The TDO technique has been previously proven to be a valuable tool \cite{Shi2019,Haravifard16,Steinhardt2019} for probing both the behavior of pure \scbo ~in the spin-dimer phase and the ambient-pressure behavior of doped \mgscbo. It allows measurement of the change in magnetization at sub-Kelvin temperatures, high pressures, and high magnetic fields (see Appendix~\ref{sec:tdo}for details). Therefore, the most valuable aspect of the technique is revealed when it is used in a sample environment that combines low temperature, high pressures and high fields. 
\begin{figure}
\centerline{\includegraphics[width=0.45\textwidth]{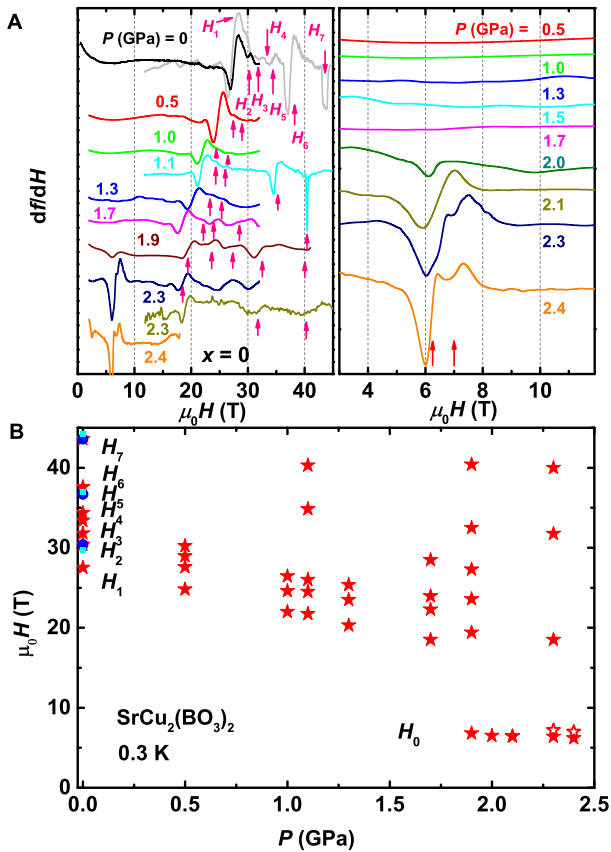}}
\caption{$P$-dependence of the magnetization plateaus and emergence of low-field anomalies in \scbo.  \textbf{A}(left panel): $df/dH$ vs $H$ for $P$ up to 2.4 GPa at 0.3 K. The data consist of results from multiple runs on different samples using a 18 T superconducting magnet, a 35 T resistive magnet, and a 45 T hybrid magnet ($H \parallel ab$ for all measurements). Red arrows denote $H_{1}$, $H_{2}$, $H_{3}$, $H_{4}$, $H_{5}$, $H_{6}$, and $H_{7}$ at $P = 0$. Here $H_{1}$, $H_{2}$, $H_{6}$, and $H_{7}$ corresponding to the sub-1/8 anomaly and the 1/8, 1/4, 1/3 plateaus, respectively. $H_{3}$, $H_{4}$, and $H_{5}$ are likely intermediate 2/15, 1/7, and 1/6 plateaus. The 1/8 plateau is identified as the shoulder that appears at a slightly higher field than the large sub-1/8 anomaly (see Supplementary Fig.~S1~\cite{SM}; for identification of the other features such as the 1/5 plateau, see Appendix~\ref{sec:exp3} and Supplementary Fig.~S2~\cite{SM}). \textbf{A}(right panel): Magnified view of the low-field behavior, showing the emergence of the low-field anomaly, which splits above $P$ $\sim$ 2.2 GPa as indicated by the two red arrows. The 2.3 GPa and 2.4 GPa traces are from measurements on two different samples using a resistive magnet and superconducting magnet. Traces in \textbf{a} and \textbf{b} are shifted vertically for clarity. \textbf{B}: $H-P$ phase diagram showing all the anomalies. $H_{1}$ to $H_{7}$ indicate the sub-1/8 anomaly and the magnetization plateaus at ambient pressure, and $H_{0}$ indicates the low-field anomaly. The red open symbols indicate the splitting of the low-field anomaly at higher $P$.}
\label{fig:Fig1}
\end{figure}

In Fig. \ref{fig:Fig1}A, we show TDO magnetic susceptibility measurements for pure \scbo ~in $\mu_{0}H$ up to 45 T ($H \parallel ab$) and $T$ = 0.3 K, where $df/dH \propto dM^{2}/d^{2}H$ (Ref. \cite{Shi2019}), for a series of pressures spanning the spin-dimer and the putative 4-spin plaquette phases. The high sensitivity of the technique allows the identification of weak magnetization changes that would otherwise be extremely difficult to detect. For example, at $P$ = 0, we identify seven anomalies in $M(H)$ at fields $H_{1} \sim$ 27.5 T, $H_{2} \sim$ 30.2 T, $H_{3} \sim$ 31.8 T, $H_{4} \sim$ 33.4 T, $H_{5} \sim$ 34.4 T, $H_{6} \sim$ 37.6 T, and $H_{7} \sim$ 43.6 T, all of which correspond to jumps or slope changes in magnetization (see Appendix~\ref{sec:exp3} for details). $H_{6}$ and $H_{7}$ can be identified immediately as the onset of the 1/4 and 1/3 magnetization plateaus~\cite{Shi2019}. When $P$ is increased to 1.1 GPa, similar anomalies are observed, but shifted to lower fields. In the intermediate plaquette phase, at 1.9 GPa and 2.3 GPa, we still can identify two anomalies in this field range, although they are now much weaker and shifted slightly to lower fields. It is tempting to assign these two anomalies as extensions of the $H_{6}$ (1/4 plateau) and $H_{7}$ (1/3 plateau) seen at 0 and 1.1 GPa. However, we caution that the fate of the magnetization plateaus at higher pressure needs to be understood first. Indeed, as we will show below, the real physical picture is much more complicated than a simple extension from the ambient pressure results. In fact, some of the anomalies actually signal entirely new phases that only appear at high pressure and high field, such as a 1/5 plateau phase and a $10 \times 2$  supersolid phase. 

The interpretation of the data obtained at lower fields also requires some care. First, at $P$ = 0, we identify three anomalies at $H_{3}$, $H_{4}$, and $H_{5}$, between the expected 1/4 and 1/8 plateaus (see Fig. \ref{fig:Fig1} and Supplementary Fig. S2A~\cite{SM}). Here, NMR measurements have found evidence for 2/15 and 1/6 plateaus~\cite{Takigawa2013} for $H \parallel c$. After accounting for the g-factor difference between $H \parallel c$ and $H \parallel ab$, we find that two of the anomalies ($H_{3}$ and $H_{5}$) are located at fields consistent with the onsets of the 2/15 and 1/6 plateaus~\cite{Takigawa2013}. For $H_{4}$, it is likely associated with the 1/7 anomaly, the possible trace of an intermediate plateau~\cite{corboz14_shastry} stabilized by factors such as inter-layer coupling. Further details of our results at ambient pressure are discussed in Appendix~\ref{sec:exp4}.

Unlike at high field, the low-field behavior in the plaquette state can be immediately seen to be qualitatively different from the spin-dimer phase, with the emergence of a new feature near 7 T (Fig.~\ref{fig:Fig1}A, right panel), which we refer to as $H_0$. This feature further splits at 2.2 GPa. It is interesting to note that the magnetic energy scale of this anomaly is comparable to the low-energy excitation mode observed by inelastic neutron scattering in the plaquette state~\cite{Zayed2017}, albeit without observing the subsequent splitting at the higher pressure. Structure factor measurements of this low-energy mode suggested that the ground state is a full plaquette featuring diagonal bonds~\cite{Zayed2017}.
As shown below, our numerical results show that the splitting corresponds to a hidden AFM state, which is possibly connected adiabatically to the AFM ground state observed by heat capacity measurements above 2.5 GPa. This is consistent with the expectation that the AFM phase is favored at higher $T$, $H$, and $P$ where entropy is increased~\cite{Jimenez2021}.

In Fig. \ref{fig:Fig1}B, we show the characteristic magnetic fields of all the anomalies as a function of pressure. Here, $H_{2}$, $H_{6}$, and $H_{7}$ correspond to the 1/8, 1/4, and 1/3 plateaus respectively at ambient pressure; $H_{1}$ is the sub-1/8 anomaly that signals the onset of the condensation of triplet bound states; $H_{3}$, $H_{4}$, and $H_{5}$ are attributed to the intermediate magnetization plateaus as discussed above; $H_{0}$ represents the low-field anomalies that appear above 1.7 GPa. As we will show below, these characteristic fields constitute a rich phase diagram containing a variety of spin superstructures. 

Finally we have also investigated the effect of Mg dopants in the system (see Appendix C) and found that the results can be consistently explained along the lines of the impurity-induced spin structures we established for ambient pressure~\cite{Shi2019}. 

We note that our low-field results are consistent with previous results in the entire pressure range. At ambient pressure, \scbo ~has a 3 meV gap separating the spin singlet ground state and the triplet excited state~\cite{Haravifard2006}. Applying pressure within the dimer phase suppresses this gap~\cite{Haravifard2012}, but it does not completely close before  entering the plaquette state \cite{Zayed2017}. Inelastic neutron scattering measurements within the plaquette phase found the emergence of a low-energy mode along with a slight hardening of the triplon gap~\cite{Zayed2017}. The pressure dependence of the former was tracked via heat capacity and was found to decrease with increasing pressure~\cite{Guo2020}. The spin gap can also be suppressed by the Zeeman mechanism, where the lowest excited state is brought down in energy by the application of the magnetic field. Therefore, we focus on the pressure dependence of the characteristic fields identified by our TDO measurements.

\begin{figure}
\centerline{\includegraphics[width=0.45\textwidth]{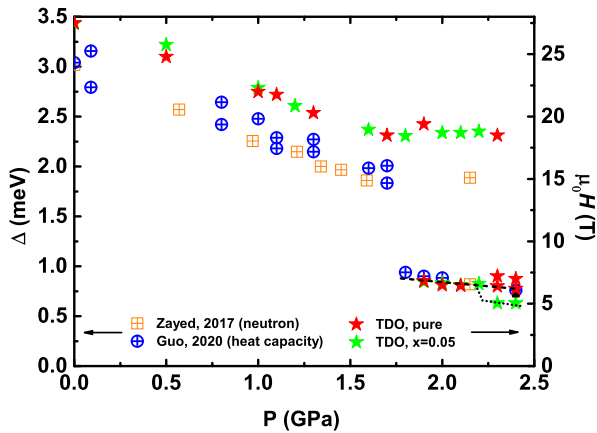}}
\caption{$H-P$ phase diagram of the sub-1/8 anomaly and the LE mode, compared with the $\Delta-P$ phase diagram established by neutron scattering~\cite{Zayed2017} and heat capacity~\cite{Guo2020}. (Left axis)  Orange squares and blue circles are the spin gap values reported by studies of neutron scattering~\cite{Zayed2017} and heat capacity~\cite{Guo2020}, respectively. (Right axis) Red and green stars are TDO results for the $x = 0$ and $x = 0.05$ samples. Similar pressure dependence are observed for $H$ \emph{vs.} $P$ and $\Delta$ \emph{vs.} $P$.}
\label{fig:Fig4}
\end{figure}

We first focus on $H_1$, i.e. the condensation field of the spin-2 bound states (see Appendix B), and its pressure dependence. 
Interestingly, as shown in  Fig.~\ref{fig:Fig4}, there is a similarity between the pressure-dependence of some of the characteristic fields ($\mu_{0}H_1$ and $\mu_{0}H_0$) and that of the spin gap measured by neutron scattering and heat capacity measurements.  However, some notable differences of the two types of pressure dependence are also observed at $P \gtrsim$ 2.3 GPa. Here, $\mu_{0}H_0$ splits, signalling the emergence of the AFM state. Finally, we note that while the introduction of Mg doping does not qualitatively change the behavior, the new modes presaging the AFM state are shifted to lower energy compared to pure \scbo.

In the spin dimer phase, adding impurities has been found not to move the onset fields of the 1/n plateaus, although increased impurity concentration does soften the spin superstructures and enhances the probability of forming impurity pairs and impurity-induced triplet states~\cite{Shi2019}. This suggests that the superstructures of the triplet bound states have excitation energies independent of impurity doping.  As shown in Fig.~\ref{fig:Fig4}, this similarity between the pure and doped cases extends into the plaquette phase, indicating that the triplon excitation is likewise insensitive to impurity doping. However, the impurity-driven shift in the low-field mode noted above suggests that the dopants act to destabilize the plaquette phase and instead favor the AFM phase. 

\section{iPEPS calculation results}

We have performed iPEPS simulations (see Appendix~\ref{sec:ipeps}) of the SS model in a magnetic field, given by the Hamiltonian: 
\begin{equation}
H=J\sum_{\langle i,j \rangle}\bm S_{i}\cdot \bm S_{j}+J'\sum_{\langle \langle i,j \rangle\rangle}\bm
S_{i}\cdot \bm S_{j}-h\sum_{i}S_i^z,
\end{equation}
where $J$ and $J'$ are the intra-dimer and inter-dimer couplings, respectively, and the strength of the external magnetic field is controlled by $h$. At ambient pressure a ratio $J'/J=0.63$  was determined from a fit to high magnetic field data~\cite{matsuda13}. 
Applying pressure leads to an effective increase of the ratio $J'/J$, however, the precise pressure dependence of $J'$ and $J$ is not  known. In a recent ab-initio study~\cite{Badrtdinov2020}, it was found that the change in $J'$ as a function of pressure is small compared to that of $J$. Here we model the pressure dependence assuming a linear dependence of $J$ and $J'$ on pressure and  a small change of $3\%$ in $J'$ between its value at ambient pressure and its value at the critical pressure $p_c=1.8$ GPa. At ambient pressure we use $J=81.5$ K. This value lies in between previously predicted values~\cite{matsuda13,wietek19} and yields good agreement with the onset of the 1/4 and 1/3 plateaus observed in experiments.

\begin{figure*}
\centerline{\includegraphics[width=1\textwidth]
{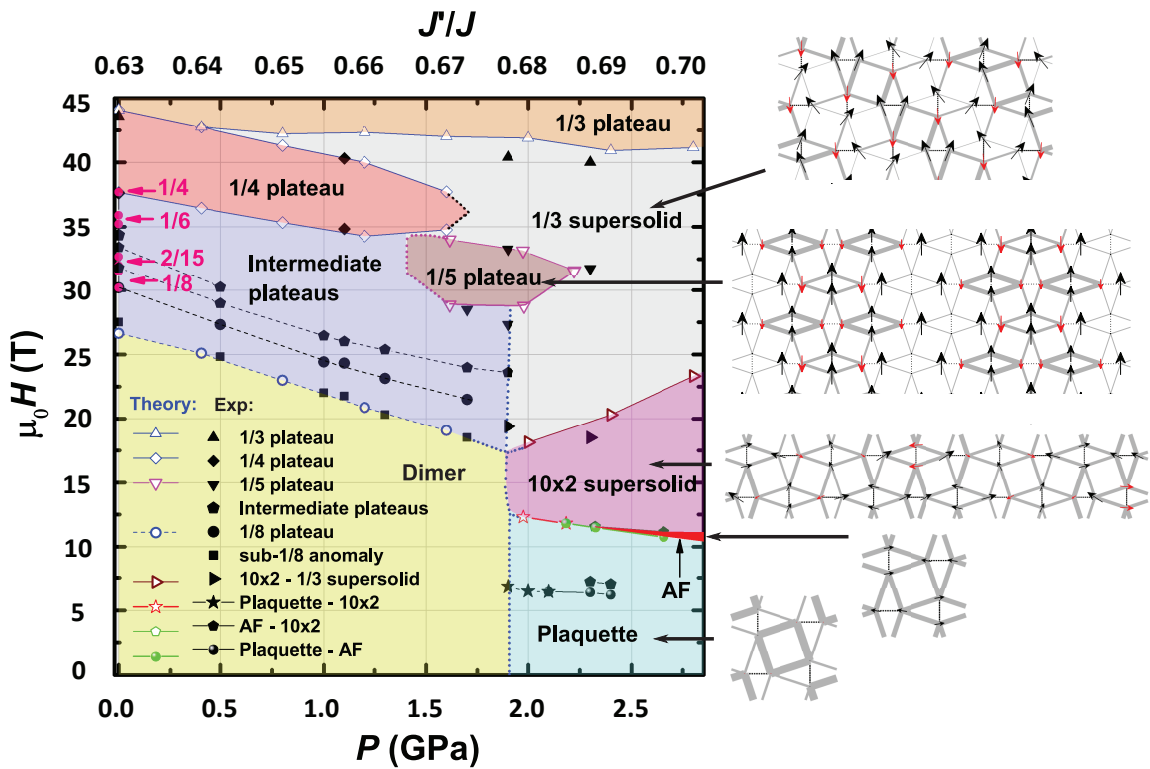}}
\caption{$H-P$ phase diagram (theory vs experiment). 
Black symbols and lines correspond to anomalies found in experiments, the colored symbols and lines are based on iPEPS results ($D=8$). The corresponding coupling ratios $J'/J$ are shown on the top axis. The colored phase regions are determined by the iPEPS data points. Dotted lines are a guide to the eye. The experimental data and the iPEPS data agree very well, except for two places: near the 1/8 plateau and the Plaquette-AFM-10x2-supersolid transitions. The iPEPS calculation does not capture the sub-1/8 anomaly which is very pronounced for $H$ $\parallel$ $ab$ but invisible for $H$ $\parallel$ $c$ (see Ref. \cite{Onizuka2000}) because of the isotropic nature of the standard SS model. On the right hand side, typical spin patterns of the phases at high pressure are drawn. The size of the spins scale with the magnitude of the local magnetic moment, where black (red) arrows point along (opposite to) the external magnetic field. The thickness of the gray bonds scales with the local bond energy (the thicker the lower the energy).}
\label{fig:phasediagram}
\end{figure*}

We focus in the following on some of the most prominent features in the phase diagram as a function of $J'/J$ (or pressure) and magnetic field. In particular, we concentrate on the phase boundaries of the magnetization plateaus and the supersolid phases at high magnetic fields, and the competing low-energy states at high pressure. The results, summarized in Fig.~\ref{fig:phasediagram}, are obtained for $D=8$ which already provides an accurate estimate of the phase boundaries (e.g. the relative error on the phase boundaries of the 1/4 and 1/3 plateaus is less than 2\% compared to the results extrapolated to the infinite $D$ limit~\cite{matsuda13}.) 

At high fields (up to 45 T) the dominant phases are the 1/4 plateau, the 1/3 plateau, and a 1/3 supersolid phase~\cite{matsuda13}. The 1/4 plateau has a finite extent up to  $J'/J=0.675(5)$ after which the intermediate field region is dominated by the 1/3 supersolid phase. The 1/3 plateau remains stable over the entire range of $J'/J$ considered here. Below the 1/4 plateau at ambient pressure there is a sequence of small magnetization plateaus (crystals of triplet bound states)~\cite{corboz14_shastry}, denoted as "intermediate plateaus" in  Fig.~\ref{fig:phasediagram}, which are stable up to $J'/J=0.675(5)$. We also add a characteristic line indicating a lower bound for the onset of the 1/8 plateau. This line is obtained by intersecting the energy of the 1/8 plateau with that of the 1/9 plateau, a plateau which is however probably unstable towards a condensate of spin-2 bound states (see above). 

At intermediate fields and approximately $J'/J=0.68$, we find  a new type of 1/5 plateau that has not been  observed  previously; and it is different from the 1/5 plateau made of localized triplet bound states appearing at smaller $J'/J$~\cite{corboz14_shastry}. The spin structure of this new plateau exhibits a stripe pattern parallel to one set of dimers, as shown in Fig.~\ref{fig:phasediagram}, where along each stripe a strong dimer triplet alternates with a pair of weaker dimer triplets.
We will discuss the physical origin of this rather unusual structure in the next section.

At ambient pressure as well as at 1.1 GPa, we find a very good agreement between the phase boundaries of the 1/4 plateau and the critical fields of the anomalies found in experiments. At 1.9 GPa and 2.3 GPa, the anomalies at 40 T are close to the iPEPS phase boundary of the 1/3 plateau, and the anomalies at 34 T are in good agreement with the upper edge of the new 1/5 plateau. At 1.7 GPa and 1.9 GPa, we also identify two anomalies consistent with the lower edge of the new 1/5 plateau. All these features are thus well captured by the standard SS model and by the simple model for the pressure dependence of $J$ and $J'$ used here. 

Finally, we turn our focus to the low-field region at high pressure. Above the empty plaquette (P) phase in zero and small fields we find a narrow partially polarized antiferromagnetic phase (AFM), and a $10 \times 2$ supersolid state, followed by the 1/3 supersolid and 1/3 plateau phases. The corresponding spin patterns are displayed in Fig.~\ref{fig:phasediagram} and examples of magnetization curves are shown in the Supplementary Figs. S7 and S8~\cite{SM}.  Note that the pattern of the $10 \times 2$ supersolid phase is different from that of the stripe phase reported in Ref.~\onlinecite{Wang2018}. Interestingly, the anomalies at 1.9 GPa and 2.3 GPa around 21.5 T in experiments lie close to the phase boundary between the $10 \times 2$ and 1/3 supersolid phase. As we shall see below, the $10 \times 2$ supersolid phase can be seen as a descendant of the unusual 1/5 plateau. 

The narrow AFM region, which vanishes around $J'/J\approx 0.686$  and which becomes broader with increasing $J'/J$, is qualitatively compatible with the splitting of the two anomalies observed at low fields in experiments. However, quantitatively we find that these phase boundaries occur at higher fields than in experiments. We believe the main reason for this discrepancy is the lack of inter-layer coupling in our model, which is of  order of $0.09J$~\cite{Miyahara03} and which is expected to enhance the stability of the AFM phase compared to the plaquette state~\cite{Miyahara03} leading to a shift of the phase boundary to smaller critical fields. Additionally, Dzyaloshinskii-Moriya (DM) interactions~\cite{Cepas01,Miyahara03,Kodama2005,romhanyi11,Haravifard2014} of order of a few percent of $J$ may affect the location of the phase boundaries. 
We note finally that the empty plaquette state is different from the full plaquette state implied by experiments, which can be obtained using a deformed Shastry-Sutherland model~\cite{moliner11,Boos2019a} that includes two types of intra- and inter-dimer interactions. These modifications of the model may also affect the magnetization process, particularly at low fields. 

\section{Nature of the 1/5 plateau and $10 \times 2$ supersolid} 

As stated in the Introduction, the high-field plateaus can be thought of as Wigner crystals of $T_1$ magnetic particles, while the lower plateaus are better interpreted as Wigner crystals of spin-2 bound states of such $T_1$ particles~\cite{corboz14_shastry}. The resulting structures are very simple to visualize. The high-field plateaus build diagonal stripes (in a geometry where dimers are horizontal or vertical), while the Wigner crystals of spin-2 bound states consists in putting the bound states as far as possible from each other. 

The two new phases discovered in the present paper, the 1/5 plateau and the $10 \times 2$ supersolid, are completely different and cannot be understood in these terms. Since the $10 \times 2$ supersolid can be understood as its descendant, let us first concentrate on the 1/5 plateau. Its main properties are: i) The stripes are vertical, and not diagonal; ii) The state is not a simple Wigner crystal of $T_1$ particles, but half the magnetic particles are delocalized over two dimers; iii) This gain of kinetic energy cannot be achieved with the mechanism that explains the other plateaus. Indeed, in order for a triplet to jump on a horizontal next-nearest neighboring dimer, this neighbor must be occupied by a non-magnetic particle called $T_0$, a triplet with zero magnetization along the field. In fact, the necessity to include $T_0$ particles in the description of this plateau has been proven in the context of a similar plateau found in a thin-tube version of the SS model made of two orthogonal dimer chains~\cite{Manmana_2011}.  

\begin{figure}
\centerline{\includegraphics[width=0.5\textwidth]{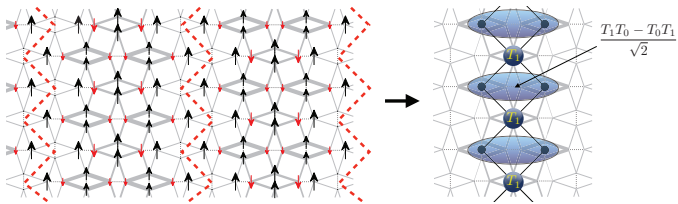}}
\caption{Nature of the new 1/5 plateau. The spin structure consists of vertical stripes of partially polarized full plaquettes which are separated by vertical singlets along the red dashed line. Each stripe can effectively be described by a spin-1 diamond chain in a magnetic field at filling 2/3, in which polarized S=1 spins ($T_1$) and dimer triplets, made of an antisymmetric combination of $T_1$ and $T_0$, are alternating.    }
\label{fig:plateau15}
\end{figure}

Interestingly, this plateau can be easily understood as a descendant of the full plaquette phase. In the full plaquette phase, the lattice symmetry is broken, and half the dimers, say the horizontal ones, are singlets, while the other ones are triplets. These triplets build effective Haldane spin-1 chains~\cite{moliner11}, leading to the full plaquette phase (called for that reason the Haldane phase in earlier papers). Similarly, in the 1/5 plateau, all horizontal dimers are singlets, but a subset of vertical dimers along vertical paths also form dimer singlets, effectively cutting the system into vertical stripes, see Fig.~\ref{fig:plateau15}. Quite remarkably, each vertical stripe is equivalent to a well known model of 1D quantum magnetism, the spin-1 diamond chain, each spin-1 corresponding to a triplet on a vertical dimer. The physics of this model is very simple to understand. It can be seen as an alternating set of single spins and dimers, and, because of the symmetric position of the dimers with respect to its neighboring single spins, the total spin of each dimer is a good quantum number. As a consequence, the magnetization process can be shown to lead to three plateaus at 1/3, 2/3 and 1 depending on whether the total spin of each dimer is a singlet (1/3), a triplet (2/3), or a quintuplet (1). The 1/5 plateau corresponds to the intermediate plateau of this diamond chain, and the wave-function of the "delocalized" particle is an antisymmetric combination of a $T_1$ and a $T_0$ on the vertical dimers over which it is delocalized, as shown on the right in Fig.~\ref{fig:plateau15}. Note that the identification of the full plaquette phase as the parent state of this plateau is further supported by the strong $J'$ bonds within the stripes, which are clearly those of full plaquettes.

Finally, the $10 \times 2$ supersolid phase can be obtained from the 1/5 plateau by an alternating rotation of the magnetization of successive stripes clockwise or counterclockwise, and by adding some magnetic particles between the stripes, see Supplementary Fig.~S9~\cite{SM}. Below we further discuss some other interesting aspects of our results beyond the 1/5 plateau and the $10 \times 2$ supersolid phases. 

\section{Discussion}
 
In summary, our results at high magnetic field and high pressure using TDO magnetization measurements and iPEPS calculations elucidate the very rich multi-dimensional ($H-T-P-x$) phase diagram of \scbo. Several phases have been reported for the first time, including an intermediate-field AFM state, and two radically new structures, a 1/5 plateau and a related $10 \times 2$ supersolid. We have argued that the latter two cannot be understood in terms of Wigner crystals of $T_1$ magnetic particles (or bound states thereof), but that they instead correspond to descendants of the full-plaquette state, which at zero field is energetically very close to the empty plaquette state in the Shastry-Sutherland model~\cite{Boos2019a}, and which is believed to be realized in \scbo. 

The physics of \scbo ~under pressure seems to be dominated by a tendency towards an orthorhombic distortion that stabilizes the full plaquette phase and its descendants. These findings make the investigation of the symmetry of the intermediate plaquette phase more relevant than ever, and suggest that the transition between this phase and the antiferromagnetic phase should be revisited in the context of a model that includes this tendency. Ideally this should be done in the context of a model where the orthorhombic distortion is spontaneously broken, but this might require including the coupling to the lattice, a formidable challenge for numerical methods. This is left for further investigation. 

Beyond \scbo, we note that the freedom to tune across energy scales that are relevant to the underlying interactions is crucial in research of strongly correlated systems, and often requires the use of extreme experimental conditions. In that respect, our experimental results provide a road-map for exploring correlated matter in extreme environments of low temperature, high magnetic field, and high pressure. Indeed, supplemented by the powerful iPEPS simulations, our results reveal the extent of the phase diagram of a prototypical correlated system that would otherwise be impossible to access. Although the TDO technique itself has been well developed, its high adaptability in different sample environments and high sensitivity to detect magnetization changes have only been exploited recently, and the present work goes one step forward by combining high field and high pressure.  Similar efforts in redefining the capability of other existing experimental techniques could be a rewarding direction in future exploration of strongly correlated matter. 

\begin{acknowledgements}

We are grateful to Casey Marjerrison for her assistance with crystal growth activities at the early stages of this project. A portion of this work was performed at the National High Magnetic Field Laboratory, which is supported by the National Science Foundation Cooperative Agreement No. DMR-1157490 and DMR-1644779, the State of Florida and the U.S. Department of Energy. Z.S., S.D., W.S., and S.H. acknowledge support provided by funding from the Powe Junior Faculty Enhancement Award, and William M. Fairbank Chair in Physics at Duke University. D.M.S. and T.F.R. acknowledge support from US Department of Energy Basic Energy Sciences Award DE-SC0014866. P.C. and F.M. acknowledge the support provided by Swiss National Science Foundation and the European Research Council (ERC) under the European Union’s Horizon 2020 research and innovation programme (grant agreement No 677061).
\end{acknowledgements}

\begin{Author Contributions}

Research conceived by S.H.; Single-crystal SCBO samples grown by  H.A.D. and S.H.; High-field measurements performed by Z.S., S.D., W.S., D.G., and S.H., and analyzed by Z.S., S.D., D.M.S., T.F.R and S.H.; iPEPS calculations performed by P.C. and F.M.; Manuscript written by Z.S., P.C., F.M., D.M.S., T.F.R. and S.H.; all authors commented on the manuscript.
 
\end{Author Contributions}

\begin{appendix}
\section{Methods}
\label{sec:methods}

\subsection{Sample synthesis and characterization.} 
\label{sec:sample}
High quality single crystal samples of both \scbo ~and \mgscbo ~($x=0.02,0.03,$ and 0.05$)$ were grown by the optical floating zone technique using self-flux, at a growth rate of $0.2~mm~h^{-1}$ in an $O_{2}$ atmosphere~\cite{Dabkowska2007,Shi2019}. 

\subsection{Tunnel diode oscillator (TDO)} 
\label{sec:tdo}
Cylinder-shaped crystals of $\sim$ 2 mm in length and $\sim$ 1 mm in diameter are used for the TDO measurements. The sample was placed inside a detection coil with inductance $L$, which is small enough to be inserted into a Copper-Beryllium piston pressure cell with a maximum pressure rating of 2.4 GPa. Together with a capacitor they form a LC circuit. The changes in sample magnetization is reflected as the change in resonance frequency, which can be measured to high precision. The experiments were conducted at the dc field facility of the National High Magnetic Field Laboratory in Tallahassee, FL. 

\subsection{Identification of the magnetization anomalies in the TDO data} 
\label{sec:exp3}
The extreme sensitivity of the TDO frequency to the magnetization change allows an accurate determination of the rich magnetic phase diagram of \scbo. At ambient pressure, the TDO anomalies have been identified as magnetization changes associated with the magnetization plateaus and other magnetic phases~\cite{Haravifard16,Shi2019}. These anomalies also evolve under pressure, and we are able to track them up to 2.4 GPa. 

Because the TDO frequency is related to the magnetization by $df$/$dH$ $\propto$ -$dM^{2}$/$d^{2}H$ (see Ref. \cite{Shi2019}), a local minimum (``dip") and maximum (``bump") in $df$/$dH$ correspond to where $dM$/$dH$ changes the fastest. In principle, two types of behaviors are usually observed at a magnetization anomaly in experiments. First, $df$/$dH$ could appear with only a ``dip", which corresponds to a slope change in $M(H)$. A typical example is shown in Fig. 1A for the 1.1 GPa trace near 40 T, which we identify as the exit of the 1/4 plateau. Except for this one example, $df$/$dH$ appear with a ``dip" followed by a ``bump". The midpoint between the ``dip" and the ``bump" corresponds to where the rate of change in $M$ is the highest, i.e. jump in magnetization. Therefore, for all the TDO anomalies, we identify the ``dips", then the ``bumps" when possible, and find their midpoint, which are used as the data points plotted in the phase diagrams. Typical examples are shown in Figs. S2 and S3~\cite{SM}. For the 1/3 plateau, however, no ``bumps" can be identified except for those at high pressures ($P > \sim$1.8 GPa), due to the limitation in available magnetic field. Therefore, we identify the exit of the 1/4 plateau and the onset of the 1/3 plateau with the field values of the "dips". 

Moreover, the TDO signal is susceptible to the background interference because of its ultra high sensitivity. Therefore, we have only identified the TDO features that are repeatable for different samples in different magnet runs. As shown in Fig. S2 and S3~\cite{SM}, the 1/5 plateau is identified for (1) two different $x$=0 samples that are measured in two different magnet runs using a resistive magnet and a hybrid magnet respectively; (2) both the $x$=0 and $x$=0.05 samples. Meanwhile, the features should also have a clear trend with the pressure. 

For $P$ = 0, we also make comparison between our TDO data and the reported magnetization data~\cite{Onizuka2000} and NMR data~\cite{Takigawa2013}. The results agree very well (see Fig.~\ref{fig:phasediagram} and Fig.~S1~\cite{SM}). Note that the NMR data in Ref.~\cite{Takigawa2013} was conducted with $H$ $\parallel$ $c$. Therefore, we have converted the field values of the magnetization plateaus to those of $H$ $\parallel$ $ab$ by multiplying them with the g-factor ratio $g_{{\parallel}c}/g_{{\parallel}ab}=$2.28/2.04=1.12 (Ref. \cite{Shi2019}). The obtained results agree very well with our iPEPS calculation, as shown in Fig. \ref{fig:phasediagram}. Examples of our detailed analyses are given in Fig. S1, S2, and S3. 

\subsection{Infinite projected entangled pair states (iPEPS)} 
\label{sec:ipeps}
An iPEPS is a variational tensor network ansatz to represent 2D ground states in the thermodynamic limit~\cite{verstraete2004,jordan2008,nishio2004}, where the accuracy is systematically controlled by the bond dimension $D$ of the tensors. The ansatz consists of a unit cell of tensors, here with one tensor per dimer~\cite{Corboz13_shastry}. The optimization of the variational parameters is  done based on an imaginary time evolution using the simple update approach~\cite{jiang2008}, which provides good estimates of ground state energies while being computationally affordable even for large unit cell sizes. For small unit cells with  two tensors we further improved the results using the fast full update optimization~\cite{jordan2008,corboz2010,phien15} and we made use of the variational optimization~\cite{corboz16b} to create initial states at low~$D$. The contraction of the infinite 2D tensor network is done by a variant~\cite{corboz2011,corboz14_tJ} of the corner-transfer matrix method~\cite{nishino1996,orus2009-1}.  Further details on the iPEPS approach can be found in Refs.~\cite{corboz2010,Corboz13_shastry,phien15}.

\section{Further results at ambient pressure} 
\label{sec:exp4}
At ambient pressure and the field expected for the beginning of the 1/8 plateau, we consistently observe the weak feature marked as $H_2$ ($H\parallel ab$) (see Fig. \ref{fig:Fig1}A and Supplementary Fig. S1~\cite{SM}). Additionally, at a slightly lower field, we find a pronounced sub-1/8 anomaly at $H_{1}$, which seems to only appear for the $H \parallel ab$ orientation, and which corresponds to the large jump in magnetization that was reported in early studies~\cite{Onizuka2000} but not studied in detail. It has been suggested that any anomalies in this field range might be a hallmark of a higher order (e.g. 1/9 or 1/10) plateau~\cite{Onizuka2000,Sebastian2008,Jaime2012}. However, differences in the magnetization processes for $H \parallel c$ and $H \parallel ab$ are not consistent with the higher order plateaus, despite the similarity of the normalized values of magnetizations in the two orientations for this field range.  

An alternative explanation is  that the pronounced anomaly at $H_1$ corresponds to the transition between the single singlet condensation and the condensation of the bound states of triplets, before they crystallize at the 1/8 plateau. 
This behavior is more pronounced for $H \parallel ab$ than for $H \parallel c$, likely because the small separation between the two field scales is more apparent with the smaller g-factor along the $a$ and $b$ axes than along $c$. 

\section{Effect of Mg dopants}
\label{sec:exp5}

\begin{figure}
\centerline{\includegraphics[width=0.5\textwidth]{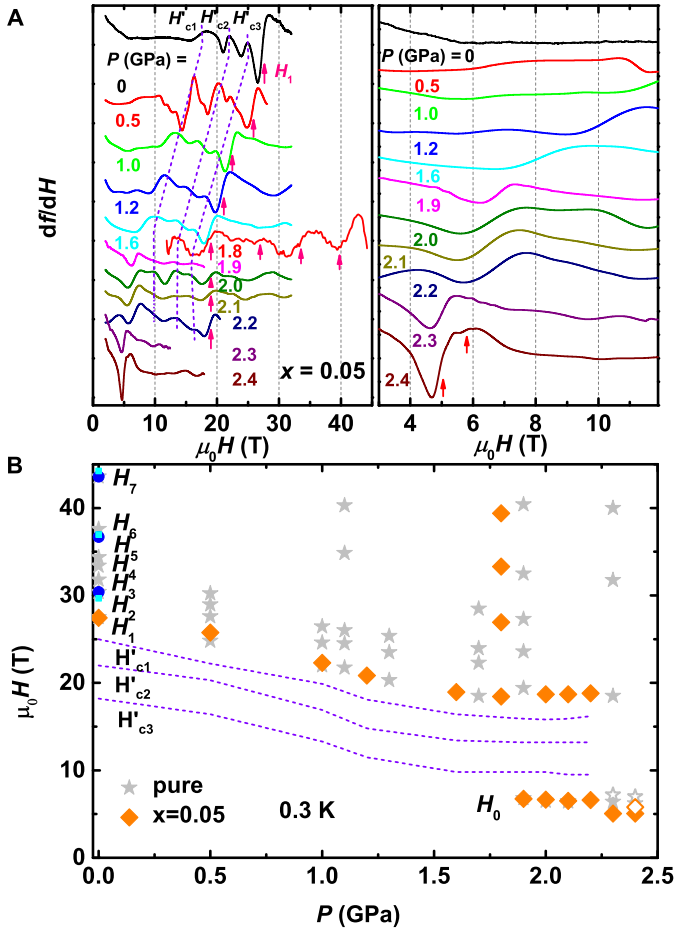}}
\caption{$P$-dependence of the magnetization plateaus and emergence of the low-field anomaly in \mgscbo, for $x=0.05$ \textbf{A}(left panel): $df/dH$ vs $H$ for $P$ up to 2.4 GPa at 0.3 K. The data consist of results from multiple runs on different samples using superconducting, resistive, and hybrid magnets ($H \parallel ab$ for all measurements). The $H_{1}$ anomaly is identified for the $P$=0 trace and tracked to higher pressure, as indicated by red arrows. Two additional anomalies at higher fields are indicated by the red arrows for the 1.8 GPa trace. The 1/8 plateau feature is much weaker and not possible to be tracked in the doped sample. The purple dashed lines show the evolution of the $H'_{c1}$, $H'_{c2}$, and $H'_{c3}$ anomalies (See Ref.~\cite{Shi2019} for a detailed study at $P$ = 0). For the 1.8 GPa trace where data is available up to 45 T, the 1/5 and 1/3 plateaus are also indicated by the red arrows (see Supplementary Fig. S3~\cite{SM} for details). \textbf{A}(right panel): Magnified view of the low-field behavior showing the pressure dependence of the low-field anomaly. A sudden softening of the mode is seen above $P$ $\sim$ 2.2 GPa, where a splitting is also observed, as indicated by the two red arrows, similar to the $x=0$ case. \textbf{B}: $H-P$ phase diagram containing the data for $x = 0.05$  (orange) in comparison with $x=0$ (grey). The open symbols indicate the splitting of the low-field anomaly at higher $P$. The purple dashed lines show the pressure dependence of the $H'_{c1}$, $H'_{c2}$, and $H'_{c3}$ anomalies.}
\label{fig:Fig2}
\end{figure}

At ambient pressure, the impurities introduced in \mgscbo ~were found to form pairs and coexist with the 2-spin dimer singlets at $H = 0$, and break down upon application of a moderate field, whereas at higher fields, the impurities interact strongly with the triplets and their bound states~\cite{Shi2019}. Four anomalies denoted by $H_{0}'$, $H_{1}'$, $H_{2}'$, $H_{3}'$ were observed at fields below the 1/8 plateau~\cite{Shi2019}. Here, $H_{0}'$ and $H_{1}'$ correspond to breakups of the impurity pairs, $H_{2}'$ corresponds to the emergence of localized bound states of triplets, and $H_{3}'$ marks the appearance of additional triplet excitations. In Fig. \ref{fig:Fig2}, we show how these features evolve with the application of pressure. We note first that when measuring inside a pressure cell, the very weak anomaly of $H_{0}'$ is not observed. However, the $H_{1}'$, $H_{2}'$ and $H_{3}'$ anomalies can be tracked clearly well above 1.7 GPa, suggesting that the impurities interact with the 4-spin plaquette in a similar manner as they do with the spin dimers. This suggests that the picture of impurity-induced spin structures we established for ambient pressure~\cite{Shi2019} also applies at high pressure. 

At higher fields, we still can identify the sub-1/8 anomaly at $H_{1}$ and track it to high pressure, though the more subtle 1/8 anomaly at $H_{2}$ is no longer visible in the doped sample. The low-field anomaly at $H_{0}$ also is observed (see Fig. \ref{fig:Fig2}A right panel), which again splits at even higher pressure $\sim$ 2.4 GPa. Notably, the field scale $H_{0}$ for the $x=0.05$ sample is reduced compared to that for the pure sample. The difference emerges clearly in the doping dependence at 2.4 GPa in Supplementary Fig.~S5~\cite{SM}, as well as in Fig. \ref{fig:Fig2}B, where the characteristic energies for $x=0.05$ are added to the phase diagram together with those for the pure sample.\\

\end{appendix}

\bibliographystyle{apsrev4-2}
\bibliography{SCBO_references.bib}

\begin{thebibliography}{59}%
\makeatletter
\providecommand \@ifxundefined [1]{%
 \@ifx{#1\undefined}
}%
\providecommand \@ifnum [1]{%
 \ifnum #1\expandafter \@firstoftwo
 \else \expandafter \@secondoftwo
 \fi
}%
\providecommand \@ifx [1]{%
 \ifx #1\expandafter \@firstoftwo
 \else \expandafter \@secondoftwo
 \fi
}%
\providecommand \natexlab [1]{#1}%
\providecommand \enquote  [1]{``#1''}%
\providecommand \bibnamefont  [1]{#1}%
\providecommand \bibfnamefont [1]{#1}%
\providecommand \citenamefont [1]{#1}%
\providecommand \href@noop [0]{\@secondoftwo}%
\providecommand \href [0]{\begingroup \@sanitize@url \@href}%
\providecommand \@href[1]{\@@startlink{#1}\@@href}%
\providecommand \@@href[1]{\endgroup#1\@@endlink}%
\providecommand \@sanitize@url [0]{\catcode `\\12\catcode `\$12\catcode
  `\&12\catcode `\#12\catcode `\^12\catcode `\_12\catcode `\%12\relax}%
\providecommand \@@startlink[1]{}%
\providecommand \@@endlink[0]{}%
\providecommand \url  [0]{\begingroup\@sanitize@url \@url }%
\providecommand \@url [1]{\endgroup\@href {#1}{\urlprefix }}%
\providecommand \urlprefix  [0]{URL }%
\providecommand \Eprint [0]{\href }%
\providecommand \doibase [0]{https://doi.org/}%
\providecommand \selectlanguage [0]{\@gobble}%
\providecommand \bibinfo  [0]{\@secondoftwo}%
\providecommand \bibfield  [0]{\@secondoftwo}%
\providecommand \translation [1]{[#1]}%
\providecommand \BibitemOpen [0]{}%
\providecommand \bibitemStop [0]{}%
\providecommand \bibitemNoStop [0]{.\EOS\space}%
\providecommand \EOS [0]{\spacefactor3000\relax}%
\providecommand \BibitemShut  [1]{\csname bibitem#1\endcsname}%
\let\auto@bib@innerbib\@empty
\bibitem [{\citenamefont {Lacroix}\ \emph {et~al.}(2011)\citenamefont
  {Lacroix}, \citenamefont {Mendels},\ and\ \citenamefont
  {Mila}}]{lacroix2011introduction}%
  \BibitemOpen
  \bibfield  {author} {\bibinfo {author} {\bibfnamefont {C.}~\bibnamefont
  {Lacroix}}, \bibinfo {author} {\bibfnamefont {P.}~\bibnamefont {Mendels}},\
  and\ \bibinfo {author} {\bibfnamefont {F.}~\bibnamefont {Mila}},\ }\href@noop
  {} {\emph {\bibinfo {title} {Introduction to frustrated magnetism: materials,
  experiments, theory}}},\ Vol.\ \bibinfo {volume} {164}\ (\bibinfo
  {publisher} {Springer Science \& Business Media},\ \bibinfo {year}
  {2011})\BibitemShut {NoStop}%
\bibitem [{\citenamefont {Kageyama}\ \emph {et~al.}(1999)\citenamefont
  {Kageyama}, \citenamefont {Yoshimura}, \citenamefont {Stern}, \citenamefont
  {Mushnikov}, \citenamefont {Onizuka}, \citenamefont {Kato}, \citenamefont
  {Kosuge}, \citenamefont {Slichter}, \citenamefont {Goto},\ and\ \citenamefont
  {Ueda}}]{Kageyama1999}%
  \BibitemOpen
  \bibfield  {author} {\bibinfo {author} {\bibfnamefont {H.}~\bibnamefont
  {Kageyama}}, \bibinfo {author} {\bibfnamefont {K.}~\bibnamefont {Yoshimura}},
  \bibinfo {author} {\bibfnamefont {R.}~\bibnamefont {Stern}}, \bibinfo
  {author} {\bibfnamefont {N.~V.}\ \bibnamefont {Mushnikov}}, \bibinfo {author}
  {\bibfnamefont {K.}~\bibnamefont {Onizuka}}, \bibinfo {author} {\bibfnamefont
  {M.}~\bibnamefont {Kato}}, \bibinfo {author} {\bibfnamefont {K.}~\bibnamefont
  {Kosuge}}, \bibinfo {author} {\bibfnamefont {C.~P.}\ \bibnamefont
  {Slichter}}, \bibinfo {author} {\bibfnamefont {T.}~\bibnamefont {Goto}},\
  and\ \bibinfo {author} {\bibfnamefont {Y.}~\bibnamefont {Ueda}},\ }\href
  {https://doi.org/10.1103/PhysRevLett.82.3168} {\bibfield  {journal} {\bibinfo
   {journal} {Phys. Rev. Lett.}\ }\textbf {\bibinfo {volume} {82}},\ \bibinfo
  {pages} {3168} (\bibinfo {year} {1999})}\BibitemShut {NoStop}%
\bibitem [{\citenamefont {Kodama}\ \emph {et~al.}(2002)\citenamefont {Kodama},
  \citenamefont {Takigawa}, \citenamefont {Horvati{\'c}}, \citenamefont
  {Berthier}, \citenamefont {Kageyama}, \citenamefont {Ueda}, \citenamefont
  {Miyahara}, \citenamefont {Becca},\ and\ \citenamefont {Mila}}]{Kodama2002}%
  \BibitemOpen
  \bibfield  {author} {\bibinfo {author} {\bibfnamefont {K.}~\bibnamefont
  {Kodama}}, \bibinfo {author} {\bibfnamefont {M.}~\bibnamefont {Takigawa}},
  \bibinfo {author} {\bibfnamefont {M.}~\bibnamefont {Horvati{\'c}}}, \bibinfo
  {author} {\bibfnamefont {C.}~\bibnamefont {Berthier}}, \bibinfo {author}
  {\bibfnamefont {H.}~\bibnamefont {Kageyama}}, \bibinfo {author}
  {\bibfnamefont {Y.}~\bibnamefont {Ueda}}, \bibinfo {author} {\bibfnamefont
  {S.}~\bibnamefont {Miyahara}}, \bibinfo {author} {\bibfnamefont
  {F.}~\bibnamefont {Becca}},\ and\ \bibinfo {author} {\bibfnamefont
  {F.}~\bibnamefont {Mila}},\ }\href {https://doi.org/10.1126/science.1075045}
  {\bibfield  {journal} {\bibinfo  {journal} {Science}\ }\textbf {\bibinfo
  {volume} {298}},\ \bibinfo {pages} {395} (\bibinfo {year}
  {2002})}\BibitemShut {NoStop}%
\bibitem [{\citenamefont {Onizuka}\ \emph {et~al.}(2000)\citenamefont
  {Onizuka}, \citenamefont {Kageyama}, \citenamefont {Narumi}, \citenamefont
  {Kindo}, \citenamefont {Ueda},\ and\ \citenamefont {Goto}}]{Onizuka2000}%
  \BibitemOpen
  \bibfield  {author} {\bibinfo {author} {\bibfnamefont {K.}~\bibnamefont
  {Onizuka}}, \bibinfo {author} {\bibfnamefont {H.}~\bibnamefont {Kageyama}},
  \bibinfo {author} {\bibfnamefont {Y.}~\bibnamefont {Narumi}}, \bibinfo
  {author} {\bibfnamefont {K.}~\bibnamefont {Kindo}}, \bibinfo {author}
  {\bibfnamefont {Y.}~\bibnamefont {Ueda}},\ and\ \bibinfo {author}
  {\bibfnamefont {T.}~\bibnamefont {Goto}},\ }\href
  {https://doi.org/10.1143/jpsj.69.1016} {\bibfield  {journal} {\bibinfo
  {journal} {J. Phys. Soc. Jpn.}\ }\textbf {\bibinfo {volume} {69}},\ \bibinfo
  {pages} {1016} (\bibinfo {year} {2000})}\BibitemShut {NoStop}%
\bibitem [{\citenamefont {Takigawa}\ \emph {et~al.}(2004)\citenamefont
  {Takigawa}, \citenamefont {Kodama}, \citenamefont {Horvati{\'c}},
  \citenamefont {Berthier}, \citenamefont {Kageyama}, \citenamefont {Ueda},
  \citenamefont {Miyahara}, \citenamefont {Becca},\ and\ \citenamefont
  {Mila}}]{takigawa04}%
  \BibitemOpen
  \bibfield  {author} {\bibinfo {author} {\bibfnamefont {M.}~\bibnamefont
  {Takigawa}}, \bibinfo {author} {\bibfnamefont {K.}~\bibnamefont {Kodama}},
  \bibinfo {author} {\bibfnamefont {M.}~\bibnamefont {Horvati{\'c}}}, \bibinfo
  {author} {\bibfnamefont {C.}~\bibnamefont {Berthier}}, \bibinfo {author}
  {\bibfnamefont {H.}~\bibnamefont {Kageyama}}, \bibinfo {author}
  {\bibfnamefont {Y.}~\bibnamefont {Ueda}}, \bibinfo {author} {\bibfnamefont
  {S.}~\bibnamefont {Miyahara}}, \bibinfo {author} {\bibfnamefont
  {F.}~\bibnamefont {Becca}},\ and\ \bibinfo {author} {\bibfnamefont
  {F.}~\bibnamefont {Mila}},\ }\href
  {https://doi.org/10.1016/j.physb.2004.01.014} {\bibfield  {journal} {\bibinfo
   {journal} {Physica B: Condensed Matter}\ }\textbf {\bibinfo {volume}
  {346{\textendash}347}},\ \bibinfo {pages} {27} (\bibinfo {year}
  {2004})}\BibitemShut {NoStop}%
\bibitem [{\citenamefont {Levy}\ \emph {et~al.}(2008)\citenamefont {Levy},
  \citenamefont {Sheikin}, \citenamefont {Berthier}, \citenamefont
  {Horvati{\'c}}, \citenamefont {Takigawa}, \citenamefont {Kageyama},
  \citenamefont {Waki},\ and\ \citenamefont {Ueda}}]{levy08}%
  \BibitemOpen
  \bibfield  {author} {\bibinfo {author} {\bibfnamefont {F.}~\bibnamefont
  {Levy}}, \bibinfo {author} {\bibfnamefont {I.}~\bibnamefont {Sheikin}},
  \bibinfo {author} {\bibfnamefont {C.}~\bibnamefont {Berthier}}, \bibinfo
  {author} {\bibfnamefont {M.}~\bibnamefont {Horvati{\'c}}}, \bibinfo {author}
  {\bibfnamefont {M.}~\bibnamefont {Takigawa}}, \bibinfo {author}
  {\bibfnamefont {H.}~\bibnamefont {Kageyama}}, \bibinfo {author}
  {\bibfnamefont {T.}~\bibnamefont {Waki}},\ and\ \bibinfo {author}
  {\bibfnamefont {Y.}~\bibnamefont {Ueda}},\ }\href
  {https://doi.org/10.1209/0295-5075/81/67004} {\bibfield  {journal} {\bibinfo
  {journal} {{EPL}}\ }\textbf {\bibinfo {volume} {81}},\ \bibinfo {pages}
  {67004} (\bibinfo {year} {2008})}\BibitemShut {NoStop}%
\bibitem [{\citenamefont {Sebastian}\ \emph {et~al.}(2008)\citenamefont
  {Sebastian}, \citenamefont {Harrison}, \citenamefont {Sengupta},
  \citenamefont {Batista}, \citenamefont {Francoual}, \citenamefont {Palm},
  \citenamefont {Murphy}, \citenamefont {Marcano}, \citenamefont {Dabkowska},\
  and\ \citenamefont {Gaulin}}]{Sebastian2008}%
  \BibitemOpen
  \bibfield  {author} {\bibinfo {author} {\bibfnamefont {S.~E.}\ \bibnamefont
  {Sebastian}}, \bibinfo {author} {\bibfnamefont {N.}~\bibnamefont {Harrison}},
  \bibinfo {author} {\bibfnamefont {P.}~\bibnamefont {Sengupta}}, \bibinfo
  {author} {\bibfnamefont {C.}~\bibnamefont {Batista}}, \bibinfo {author}
  {\bibfnamefont {S.}~\bibnamefont {Francoual}}, \bibinfo {author}
  {\bibfnamefont {E.}~\bibnamefont {Palm}}, \bibinfo {author} {\bibfnamefont
  {T.}~\bibnamefont {Murphy}}, \bibinfo {author} {\bibfnamefont
  {N.}~\bibnamefont {Marcano}}, \bibinfo {author} {\bibfnamefont
  {H.}~\bibnamefont {Dabkowska}},\ and\ \bibinfo {author} {\bibfnamefont
  {B.}~\bibnamefont {Gaulin}},\ }\href@noop {} {\bibfield  {journal} {\bibinfo
  {journal} {Proceedings of the National Academy of Sciences}\ }\textbf
  {\bibinfo {volume} {105}},\ \bibinfo {pages} {20157} (\bibinfo {year}
  {2008})}\BibitemShut {NoStop}%
\bibitem [{\citenamefont {Jaime}\ \emph {et~al.}(2012)\citenamefont {Jaime},
  \citenamefont {Daou}, \citenamefont {Crooker}, \citenamefont {Weickert},
  \citenamefont {Uchida}, \citenamefont {Feiguin}, \citenamefont {Batista},
  \citenamefont {Dabkowska},\ and\ \citenamefont {Gaulin}}]{Jaime2012}%
  \BibitemOpen
  \bibfield  {author} {\bibinfo {author} {\bibfnamefont {M.}~\bibnamefont
  {Jaime}}, \bibinfo {author} {\bibfnamefont {R.}~\bibnamefont {Daou}},
  \bibinfo {author} {\bibfnamefont {S.~A.}\ \bibnamefont {Crooker}}, \bibinfo
  {author} {\bibfnamefont {F.}~\bibnamefont {Weickert}}, \bibinfo {author}
  {\bibfnamefont {A.}~\bibnamefont {Uchida}}, \bibinfo {author} {\bibfnamefont
  {A.~E.}\ \bibnamefont {Feiguin}}, \bibinfo {author} {\bibfnamefont {C.~D.}\
  \bibnamefont {Batista}}, \bibinfo {author} {\bibfnamefont {H.~A.}\
  \bibnamefont {Dabkowska}},\ and\ \bibinfo {author} {\bibfnamefont {B.~D.}\
  \bibnamefont {Gaulin}},\ }\href@noop {} {\bibfield  {journal} {\bibinfo
  {journal} {Proceedings of the National Academy of Sciences}\ }\textbf
  {\bibinfo {volume} {109}},\ \bibinfo {pages} {12404} (\bibinfo {year}
  {2012})}\BibitemShut {NoStop}%
\bibitem [{\citenamefont {Takigawa}\ \emph {et~al.}(2013)\citenamefont
  {Takigawa}, \citenamefont {Horvatic}, \citenamefont {Waki}, \citenamefont
  {Kr\"amer}, \citenamefont {Berthier}, \citenamefont {L\'evy-Bertrand},
  \citenamefont {Sheikin}, \citenamefont {Kageyama}, \citenamefont {Ueda},\
  and\ \citenamefont {Mila}}]{Takigawa2013}%
  \BibitemOpen
  \bibfield  {author} {\bibinfo {author} {\bibfnamefont {M.}~\bibnamefont
  {Takigawa}}, \bibinfo {author} {\bibfnamefont {M.}~\bibnamefont {Horvatic}},
  \bibinfo {author} {\bibfnamefont {T.}~\bibnamefont {Waki}}, \bibinfo {author}
  {\bibfnamefont {S.}~\bibnamefont {Kr\"amer}}, \bibinfo {author}
  {\bibfnamefont {C.}~\bibnamefont {Berthier}}, \bibinfo {author}
  {\bibfnamefont {F.}~\bibnamefont {L\'evy-Bertrand}}, \bibinfo {author}
  {\bibfnamefont {I.}~\bibnamefont {Sheikin}}, \bibinfo {author} {\bibfnamefont
  {H.}~\bibnamefont {Kageyama}}, \bibinfo {author} {\bibfnamefont
  {Y.}~\bibnamefont {Ueda}},\ and\ \bibinfo {author} {\bibfnamefont
  {F.}~\bibnamefont {Mila}},\ }\href
  {https://doi.org/10.1103/PhysRevLett.110.067210} {\bibfield  {journal}
  {\bibinfo  {journal} {Phys. Rev. Lett.}\ }\textbf {\bibinfo {volume} {110}},\
  \bibinfo {pages} {067210} (\bibinfo {year} {2013})}\BibitemShut {NoStop}%
\bibitem [{\citenamefont {Matsuda}\ \emph {et~al.}(2013)\citenamefont
  {Matsuda}, \citenamefont {Abe}, \citenamefont {Takeyama}, \citenamefont
  {Kageyama}, \citenamefont {Corboz}, \citenamefont {Honecker}, \citenamefont
  {Manmana}, \citenamefont {Foltin}, \citenamefont {Schmidt},\ and\
  \citenamefont {Mila}}]{matsuda13}%
  \BibitemOpen
  \bibfield  {author} {\bibinfo {author} {\bibfnamefont {Y.~H.}\ \bibnamefont
  {Matsuda}}, \bibinfo {author} {\bibfnamefont {N.}~\bibnamefont {Abe}},
  \bibinfo {author} {\bibfnamefont {S.}~\bibnamefont {Takeyama}}, \bibinfo
  {author} {\bibfnamefont {H.}~\bibnamefont {Kageyama}}, \bibinfo {author}
  {\bibfnamefont {P.}~\bibnamefont {Corboz}}, \bibinfo {author} {\bibfnamefont
  {A.}~\bibnamefont {Honecker}}, \bibinfo {author} {\bibfnamefont {S.~R.}\
  \bibnamefont {Manmana}}, \bibinfo {author} {\bibfnamefont {G.~R.}\
  \bibnamefont {Foltin}}, \bibinfo {author} {\bibfnamefont {K.~P.}\
  \bibnamefont {Schmidt}},\ and\ \bibinfo {author} {\bibfnamefont
  {F.}~\bibnamefont {Mila}},\ }\href
  {https://doi.org/10.1103/PhysRevLett.111.137204} {\bibfield  {journal}
  {\bibinfo  {journal} {Phys. Rev. Lett.}\ }\textbf {\bibinfo {volume} {111}},\
  \bibinfo {pages} {137204} (\bibinfo {year} {2013})}\BibitemShut {NoStop}%
\bibitem [{\citenamefont {Haravifard}\ \emph {et~al.}(2016)\citenamefont
  {Haravifard}, \citenamefont {Graf}, \citenamefont {Feiguin}, \citenamefont
  {Batista}, \citenamefont {Lang}, \citenamefont {Silevitch}, \citenamefont
  {Srajer}, \citenamefont {Gaulin}, \citenamefont {Dabkowska},\ and\
  \citenamefont {Rosenbaum}}]{Haravifard16}%
  \BibitemOpen
  \bibfield  {author} {\bibinfo {author} {\bibfnamefont {S.}~\bibnamefont
  {Haravifard}}, \bibinfo {author} {\bibfnamefont {D.}~\bibnamefont {Graf}},
  \bibinfo {author} {\bibfnamefont {A.~E.}\ \bibnamefont {Feiguin}}, \bibinfo
  {author} {\bibfnamefont {C.~D.}\ \bibnamefont {Batista}}, \bibinfo {author}
  {\bibfnamefont {J.~C.}\ \bibnamefont {Lang}}, \bibinfo {author}
  {\bibfnamefont {D.~M.}\ \bibnamefont {Silevitch}}, \bibinfo {author}
  {\bibfnamefont {G.}~\bibnamefont {Srajer}}, \bibinfo {author} {\bibfnamefont
  {B.~D.}\ \bibnamefont {Gaulin}}, \bibinfo {author} {\bibfnamefont {H.~A.}\
  \bibnamefont {Dabkowska}},\ and\ \bibinfo {author} {\bibfnamefont {T.~F.}\
  \bibnamefont {Rosenbaum}},\ }\href {http://dx.doi.org/10.1038/ncomms11956}
  {\bibfield  {journal} {\bibinfo  {journal} {Nature Communications}\ }\textbf
  {\bibinfo {volume} {7}},\ \bibinfo {pages} {11956} (\bibinfo {year}
  {2016})}\BibitemShut {NoStop}%
\bibitem [{\citenamefont {Corboz}\ and\ \citenamefont
  {Mila}(2014)}]{corboz14_shastry}%
  \BibitemOpen
  \bibfield  {author} {\bibinfo {author} {\bibfnamefont {P.}~\bibnamefont
  {Corboz}}\ and\ \bibinfo {author} {\bibfnamefont {F.}~\bibnamefont {Mila}},\
  }\href {https://doi.org/10.1103/PhysRevLett.112.147203} {\bibfield  {journal}
  {\bibinfo  {journal} {Phys. Rev. Lett.}\ }\textbf {\bibinfo {volume} {112}},\
  \bibinfo {pages} {147203} (\bibinfo {year} {2014})}\BibitemShut {NoStop}%
\bibitem [{\citenamefont {C\'epas}\ \emph {et~al.}(2001)\citenamefont
  {C\'epas}, \citenamefont {Kakurai}, \citenamefont {Regnault}, \citenamefont
  {Ziman}, \citenamefont {Boucher}, \citenamefont {Aso}, \citenamefont {Nishi},
  \citenamefont {Kageyama},\ and\ \citenamefont {Ueda}}]{Cepas01}%
  \BibitemOpen
  \bibfield  {author} {\bibinfo {author} {\bibfnamefont {O.}~\bibnamefont
  {C\'epas}}, \bibinfo {author} {\bibfnamefont {K.}~\bibnamefont {Kakurai}},
  \bibinfo {author} {\bibfnamefont {L.~P.}\ \bibnamefont {Regnault}}, \bibinfo
  {author} {\bibfnamefont {T.}~\bibnamefont {Ziman}}, \bibinfo {author}
  {\bibfnamefont {J.~P.}\ \bibnamefont {Boucher}}, \bibinfo {author}
  {\bibfnamefont {N.}~\bibnamefont {Aso}}, \bibinfo {author} {\bibfnamefont
  {M.}~\bibnamefont {Nishi}}, \bibinfo {author} {\bibfnamefont
  {H.}~\bibnamefont {Kageyama}},\ and\ \bibinfo {author} {\bibfnamefont
  {Y.}~\bibnamefont {Ueda}},\ }\href
  {https://doi.org/10.1103/PhysRevLett.87.167205} {\bibfield  {journal}
  {\bibinfo  {journal} {Phys. Rev. Lett.}\ }\textbf {\bibinfo {volume} {87}},\
  \bibinfo {pages} {167205} (\bibinfo {year} {2001})}\BibitemShut {NoStop}%
\bibitem [{\citenamefont {Miyahara}\ and\ \citenamefont
  {Ueda}(2003)}]{Miyahara03}%
  \BibitemOpen
  \bibfield  {author} {\bibinfo {author} {\bibfnamefont {S.}~\bibnamefont
  {Miyahara}}\ and\ \bibinfo {author} {\bibfnamefont {K.}~\bibnamefont
  {Ueda}},\ }\href {http://stacks.iop.org/0953-8984/15/i=9/a=201} {\bibfield
  {journal} {\bibinfo  {journal} {Journal of Physics: Condensed Matter}\
  }\textbf {\bibinfo {volume} {15}},\ \bibinfo {pages} {R327} (\bibinfo {year}
  {2003})}\BibitemShut {NoStop}%
\bibitem [{\citenamefont {Kodama}\ \emph {et~al.}(2005)\citenamefont {Kodama},
  \citenamefont {Miyahara}, \citenamefont {Takigawa}, \citenamefont
  {Horvati{\'{c}}}, \citenamefont {Berthier}, \citenamefont {Mila},
  \citenamefont {Kageyama},\ and\ \citenamefont {Ueda}}]{Kodama2005}%
  \BibitemOpen
  \bibfield  {author} {\bibinfo {author} {\bibfnamefont {K.}~\bibnamefont
  {Kodama}}, \bibinfo {author} {\bibfnamefont {S.}~\bibnamefont {Miyahara}},
  \bibinfo {author} {\bibfnamefont {M.}~\bibnamefont {Takigawa}}, \bibinfo
  {author} {\bibfnamefont {M.}~\bibnamefont {Horvati{\'{c}}}}, \bibinfo
  {author} {\bibfnamefont {C.}~\bibnamefont {Berthier}}, \bibinfo {author}
  {\bibfnamefont {F.}~\bibnamefont {Mila}}, \bibinfo {author} {\bibfnamefont
  {H.}~\bibnamefont {Kageyama}},\ and\ \bibinfo {author} {\bibfnamefont
  {Y.}~\bibnamefont {Ueda}},\ }\href
  {https://doi.org/10.1088/0953-8984/17/4/l02} {\bibfield  {journal} {\bibinfo
  {journal} {Journal of Physics: Condensed Matter}\ }\textbf {\bibinfo {volume}
  {17}},\ \bibinfo {pages} {L61} (\bibinfo {year} {2005})}\BibitemShut
  {NoStop}%
\bibitem [{\citenamefont {Haravifard}\ \emph {et~al.}(2014)\citenamefont
  {Haravifard}, \citenamefont {Banerjee}, \citenamefont {van Wezel},
  \citenamefont {Silevitch}, \citenamefont {dos Santos}, \citenamefont {Lang},
  \citenamefont {Kermarrec}, \citenamefont {Srajer}, \citenamefont {Gaulin},
  \citenamefont {Molaison} \emph {et~al.}}]{Haravifard2014}%
  \BibitemOpen
  \bibfield  {author} {\bibinfo {author} {\bibfnamefont {S.}~\bibnamefont
  {Haravifard}}, \bibinfo {author} {\bibfnamefont {A.}~\bibnamefont
  {Banerjee}}, \bibinfo {author} {\bibfnamefont {J.}~\bibnamefont {van Wezel}},
  \bibinfo {author} {\bibfnamefont {D.}~\bibnamefont {Silevitch}}, \bibinfo
  {author} {\bibfnamefont {A.}~\bibnamefont {dos Santos}}, \bibinfo {author}
  {\bibfnamefont {J.}~\bibnamefont {Lang}}, \bibinfo {author} {\bibfnamefont
  {E.}~\bibnamefont {Kermarrec}}, \bibinfo {author} {\bibfnamefont
  {G.}~\bibnamefont {Srajer}}, \bibinfo {author} {\bibfnamefont
  {B.}~\bibnamefont {Gaulin}}, \bibinfo {author} {\bibfnamefont
  {J.}~\bibnamefont {Molaison}}, \emph {et~al.},\ }\href@noop {} {\bibfield
  {journal} {\bibinfo  {journal} {Proceedings of the National Academy of
  Sciences}\ }\textbf {\bibinfo {volume} {111}},\ \bibinfo {pages} {14372}
  (\bibinfo {year} {2014})}\BibitemShut {NoStop}%
\bibitem [{\citenamefont {Romh\'anyi}\ \emph {et~al.}(2011)\citenamefont
  {Romh\'anyi}, \citenamefont {Totsuka},\ and\ \citenamefont
  {Penc}}]{romhanyi11}%
  \BibitemOpen
  \bibfield  {author} {\bibinfo {author} {\bibfnamefont {J.}~\bibnamefont
  {Romh\'anyi}}, \bibinfo {author} {\bibfnamefont {K.}~\bibnamefont
  {Totsuka}},\ and\ \bibinfo {author} {\bibfnamefont {K.}~\bibnamefont
  {Penc}},\ }\href {https://doi.org/10.1103/PhysRevB.83.024413} {\bibfield
  {journal} {\bibinfo  {journal} {Phys. Rev. B}\ }\textbf {\bibinfo {volume}
  {83}},\ \bibinfo {pages} {024413} (\bibinfo {year} {2011})}\BibitemShut
  {NoStop}%
\bibitem [{\citenamefont {Romh{\'a}nyi}\ \emph {et~al.}(2015)\citenamefont
  {Romh{\'a}nyi}, \citenamefont {Penc},\ and\ \citenamefont
  {Ganesh}}]{Romhanyi2015}%
  \BibitemOpen
  \bibfield  {author} {\bibinfo {author} {\bibfnamefont {J.}~\bibnamefont
  {Romh{\'a}nyi}}, \bibinfo {author} {\bibfnamefont {K.}~\bibnamefont {Penc}},\
  and\ \bibinfo {author} {\bibfnamefont {R.}~\bibnamefont {Ganesh}},\ }\href
  {https://doi.org/10.1038/ncomms7805} {\bibfield  {journal} {\bibinfo
  {journal} {Nat Commun}\ }\textbf {\bibinfo {volume} {6}},\ \bibinfo {pages}
  {1} (\bibinfo {year} {2015})},\ \bibinfo {note} {number: 1 Publisher: Nature
  Publishing Group}\BibitemShut {NoStop}%
\bibitem [{\citenamefont {Waki}\ \emph {et~al.}(2007)\citenamefont {Waki},
  \citenamefont {Arai}, \citenamefont {Takigawa}, \citenamefont {Saiga},
  \citenamefont {Uwatoko}, \citenamefont {Kageyama},\ and\ \citenamefont
  {Ueda}}]{Waki2007}%
  \BibitemOpen
  \bibfield  {author} {\bibinfo {author} {\bibfnamefont {T.}~\bibnamefont
  {Waki}}, \bibinfo {author} {\bibfnamefont {K.}~\bibnamefont {Arai}}, \bibinfo
  {author} {\bibfnamefont {M.}~\bibnamefont {Takigawa}}, \bibinfo {author}
  {\bibfnamefont {Y.}~\bibnamefont {Saiga}}, \bibinfo {author} {\bibfnamefont
  {Y.}~\bibnamefont {Uwatoko}}, \bibinfo {author} {\bibfnamefont
  {H.}~\bibnamefont {Kageyama}},\ and\ \bibinfo {author} {\bibfnamefont
  {Y.}~\bibnamefont {Ueda}},\ }\href@noop {} {\bibfield  {journal} {\bibinfo
  {journal} {Journal of the Physical Society of Japan}\ }\textbf {\bibinfo
  {volume} {76}},\ \bibinfo {pages} {073710} (\bibinfo {year}
  {2007})}\BibitemShut {NoStop}%
\bibitem [{\citenamefont {Haravifard}\ \emph {et~al.}(2012)\citenamefont
  {Haravifard}, \citenamefont {Banerjee}, \citenamefont {Lang}, \citenamefont
  {Srajer}, \citenamefont {Silevitch}, \citenamefont {Gaulin}, \citenamefont
  {Dabkowska},\ and\ \citenamefont {Rosenbaum}}]{Haravifard2012}%
  \BibitemOpen
  \bibfield  {author} {\bibinfo {author} {\bibfnamefont {S.}~\bibnamefont
  {Haravifard}}, \bibinfo {author} {\bibfnamefont {A.}~\bibnamefont
  {Banerjee}}, \bibinfo {author} {\bibfnamefont {J.}~\bibnamefont {Lang}},
  \bibinfo {author} {\bibfnamefont {G.}~\bibnamefont {Srajer}}, \bibinfo
  {author} {\bibfnamefont {D.}~\bibnamefont {Silevitch}}, \bibinfo {author}
  {\bibfnamefont {B.}~\bibnamefont {Gaulin}}, \bibinfo {author} {\bibfnamefont
  {H.}~\bibnamefont {Dabkowska}},\ and\ \bibinfo {author} {\bibfnamefont
  {T.}~\bibnamefont {Rosenbaum}},\ }\href@noop {} {\bibfield  {journal}
  {\bibinfo  {journal} {Proceedings of the National Academy of Sciences}\
  }\textbf {\bibinfo {volume} {109}},\ \bibinfo {pages} {2286} (\bibinfo {year}
  {2012})}\BibitemShut {NoStop}%
\bibitem [{\citenamefont {Zayed}\ \emph {et~al.}(2017)\citenamefont {Zayed},
  \citenamefont {R{\"u}egg}, \citenamefont {L{\"a}uchli}, \citenamefont
  {Panagopoulos}, \citenamefont {Saxena}, \citenamefont {Ellerby},
  \citenamefont {McMorrow}, \citenamefont {Str{\"a}ssle}, \citenamefont
  {Klotz}, \citenamefont {Hamel} \emph {et~al.}}]{Zayed2017}%
  \BibitemOpen
  \bibfield  {author} {\bibinfo {author} {\bibfnamefont {M.}~\bibnamefont
  {Zayed}}, \bibinfo {author} {\bibfnamefont {C.}~\bibnamefont {R{\"u}egg}},
  \bibinfo {author} {\bibfnamefont {A.}~\bibnamefont {L{\"a}uchli}}, \bibinfo
  {author} {\bibfnamefont {C.}~\bibnamefont {Panagopoulos}}, \bibinfo {author}
  {\bibfnamefont {S.}~\bibnamefont {Saxena}}, \bibinfo {author} {\bibfnamefont
  {M.}~\bibnamefont {Ellerby}}, \bibinfo {author} {\bibfnamefont
  {D.}~\bibnamefont {McMorrow}}, \bibinfo {author} {\bibfnamefont
  {T.}~\bibnamefont {Str{\"a}ssle}}, \bibinfo {author} {\bibfnamefont
  {S.}~\bibnamefont {Klotz}}, \bibinfo {author} {\bibfnamefont
  {G.}~\bibnamefont {Hamel}}, \emph {et~al.},\ }\href@noop {} {\bibfield
  {journal} {\bibinfo  {journal} {Nature physics}\ }\textbf {\bibinfo {volume}
  {13}},\ \bibinfo {pages} {962} (\bibinfo {year} {2017})}\BibitemShut
  {NoStop}%
\bibitem [{\citenamefont {Sakurai}\ \emph {et~al.}(2018)\citenamefont
  {Sakurai}, \citenamefont {Hirao}, \citenamefont {Hijii}, \citenamefont
  {Okubo}, \citenamefont {Ohta}, \citenamefont {Uwatoko}, \citenamefont
  {Kudo},\ and\ \citenamefont {Koike}}]{Sakurai2018}%
  \BibitemOpen
  \bibfield  {author} {\bibinfo {author} {\bibfnamefont {T.}~\bibnamefont
  {Sakurai}}, \bibinfo {author} {\bibfnamefont {Y.}~\bibnamefont {Hirao}},
  \bibinfo {author} {\bibfnamefont {K.}~\bibnamefont {Hijii}}, \bibinfo
  {author} {\bibfnamefont {S.}~\bibnamefont {Okubo}}, \bibinfo {author}
  {\bibfnamefont {H.}~\bibnamefont {Ohta}}, \bibinfo {author} {\bibfnamefont
  {Y.}~\bibnamefont {Uwatoko}}, \bibinfo {author} {\bibfnamefont
  {K.}~\bibnamefont {Kudo}},\ and\ \bibinfo {author} {\bibfnamefont
  {Y.}~\bibnamefont {Koike}},\ }\href {https://doi.org/10.7566/jpsj.87.033701}
  {\bibfield  {journal} {\bibinfo  {journal} {J. Phys. Soc. Jpn.}\ }\textbf
  {\bibinfo {volume} {87}},\ \bibinfo {pages} {033701} (\bibinfo {year}
  {2018})}\BibitemShut {NoStop}%
\bibitem [{\citenamefont {Guo}\ \emph {et~al.}(2020)\citenamefont {Guo},
  \citenamefont {Sun}, \citenamefont {Zhao}, \citenamefont {Wang},
  \citenamefont {Hong}, \citenamefont {Sidorov}, \citenamefont {Ma},
  \citenamefont {Wu}, \citenamefont {Li}, \citenamefont {Meng}, \citenamefont
  {Sandvik},\ and\ \citenamefont {Sun}}]{Guo2020}%
  \BibitemOpen
  \bibfield  {author} {\bibinfo {author} {\bibfnamefont {J.}~\bibnamefont
  {Guo}}, \bibinfo {author} {\bibfnamefont {G.}~\bibnamefont {Sun}}, \bibinfo
  {author} {\bibfnamefont {B.}~\bibnamefont {Zhao}}, \bibinfo {author}
  {\bibfnamefont {L.}~\bibnamefont {Wang}}, \bibinfo {author} {\bibfnamefont
  {W.}~\bibnamefont {Hong}}, \bibinfo {author} {\bibfnamefont {V.~A.}\
  \bibnamefont {Sidorov}}, \bibinfo {author} {\bibfnamefont {N.}~\bibnamefont
  {Ma}}, \bibinfo {author} {\bibfnamefont {Q.}~\bibnamefont {Wu}}, \bibinfo
  {author} {\bibfnamefont {S.}~\bibnamefont {Li}}, \bibinfo {author}
  {\bibfnamefont {Z.~Y.}\ \bibnamefont {Meng}}, \bibinfo {author}
  {\bibfnamefont {A.~W.}\ \bibnamefont {Sandvik}},\ and\ \bibinfo {author}
  {\bibfnamefont {L.}~\bibnamefont {Sun}},\ }\href
  {https://doi.org/10.1103/PhysRevLett.124.206602} {\bibfield  {journal}
  {\bibinfo  {journal} {Phys. Rev. Lett.}\ }\textbf {\bibinfo {volume} {124}},\
  \bibinfo {pages} {206602} (\bibinfo {year} {2020})}\BibitemShut {NoStop}%
\bibitem [{\citenamefont {Boos}\ \emph {et~al.}(2019)\citenamefont {Boos},
  \citenamefont {Crone}, \citenamefont {Niesen}, \citenamefont {Corboz},
  \citenamefont {Schmidt},\ and\ \citenamefont {Mila}}]{Boos2019a}%
  \BibitemOpen
  \bibfield  {author} {\bibinfo {author} {\bibfnamefont {C.}~\bibnamefont
  {Boos}}, \bibinfo {author} {\bibfnamefont {S.~P.~G.}\ \bibnamefont {Crone}},
  \bibinfo {author} {\bibfnamefont {I.~A.}\ \bibnamefont {Niesen}}, \bibinfo
  {author} {\bibfnamefont {P.}~\bibnamefont {Corboz}}, \bibinfo {author}
  {\bibfnamefont {K.~P.}\ \bibnamefont {Schmidt}},\ and\ \bibinfo {author}
  {\bibfnamefont {F.}~\bibnamefont {Mila}},\ }\href
  {https://doi.org/10.1103/PhysRevB.100.140413} {\bibfield  {journal} {\bibinfo
   {journal} {Phys. Rev. B}\ }\textbf {\bibinfo {volume} {100}},\ \bibinfo
  {pages} {140413} (\bibinfo {year} {2019})}\BibitemShut {NoStop}%
\bibitem [{\citenamefont {Badrtdinov}\ \emph {et~al.}(2020)\citenamefont
  {Badrtdinov}, \citenamefont {Tsirlin}, \citenamefont {Mazurenko},\ and\
  \citenamefont {Mila}}]{Badrtdinov2020}%
  \BibitemOpen
  \bibfield  {author} {\bibinfo {author} {\bibfnamefont {D.~I.}\ \bibnamefont
  {Badrtdinov}}, \bibinfo {author} {\bibfnamefont {A.~A.}\ \bibnamefont
  {Tsirlin}}, \bibinfo {author} {\bibfnamefont {V.~V.}\ \bibnamefont
  {Mazurenko}},\ and\ \bibinfo {author} {\bibfnamefont {F.}~\bibnamefont
  {Mila}},\ }\href {https://doi.org/10.1103/PhysRevB.101.224424} {\bibfield
  {journal} {\bibinfo  {journal} {Phys. Rev. B}\ }\textbf {\bibinfo {volume}
  {101}},\ \bibinfo {pages} {224424} (\bibinfo {year} {2020})}\BibitemShut
  {NoStop}%
\bibitem [{\citenamefont {Jim{\'e}nez}\ \emph {et~al.}(2021)\citenamefont
  {Jim{\'e}nez}, \citenamefont {Crone}, \citenamefont {Fogh}, \citenamefont
  {Zayed}, \citenamefont {Lortz}, \citenamefont {Pomjakushina}, \citenamefont
  {Conder}, \citenamefont {L{\"a}uchli}, \citenamefont {Weber}, \citenamefont
  {Wessel}, \citenamefont {Honecker}, \citenamefont {Normand}, \citenamefont
  {R{\"u}egg}, \citenamefont {Corboz}, \citenamefont {R{\o}nnow},\ and\
  \citenamefont {Mila}}]{Jimenez2021}%
  \BibitemOpen
  \bibfield  {author} {\bibinfo {author} {\bibfnamefont {J.~L.}\ \bibnamefont
  {Jim{\'e}nez}}, \bibinfo {author} {\bibfnamefont {S.~P.~G.}\ \bibnamefont
  {Crone}}, \bibinfo {author} {\bibfnamefont {E.}~\bibnamefont {Fogh}},
  \bibinfo {author} {\bibfnamefont {M.~E.}\ \bibnamefont {Zayed}}, \bibinfo
  {author} {\bibfnamefont {R.}~\bibnamefont {Lortz}}, \bibinfo {author}
  {\bibfnamefont {E.}~\bibnamefont {Pomjakushina}}, \bibinfo {author}
  {\bibfnamefont {K.}~\bibnamefont {Conder}}, \bibinfo {author} {\bibfnamefont
  {A.~M.}\ \bibnamefont {L{\"a}uchli}}, \bibinfo {author} {\bibfnamefont
  {L.}~\bibnamefont {Weber}}, \bibinfo {author} {\bibfnamefont
  {S.}~\bibnamefont {Wessel}}, \bibinfo {author} {\bibfnamefont
  {A.}~\bibnamefont {Honecker}}, \bibinfo {author} {\bibfnamefont
  {B.}~\bibnamefont {Normand}}, \bibinfo {author} {\bibfnamefont
  {C.}~\bibnamefont {R{\"u}egg}}, \bibinfo {author} {\bibfnamefont
  {P.}~\bibnamefont {Corboz}}, \bibinfo {author} {\bibfnamefont {H.~M.}\
  \bibnamefont {R{\o}nnow}},\ and\ \bibinfo {author} {\bibfnamefont
  {F.}~\bibnamefont {Mila}},\ }\href
  {https://doi.org/10.1038/s41586-021-03411-8} {\bibfield  {journal} {\bibinfo
  {journal} {Nature}\ }\textbf {\bibinfo {volume} {592}},\ \bibinfo {pages}
  {370} (\bibinfo {year} {2021})}\BibitemShut {NoStop}%
\bibitem [{\citenamefont {Sriram~Shastry}\ and\ \citenamefont
  {Sutherland}(1981)}]{Shastry81}%
  \BibitemOpen
  \bibfield  {author} {\bibinfo {author} {\bibfnamefont {B.}~\bibnamefont
  {Sriram~Shastry}}\ and\ \bibinfo {author} {\bibfnamefont {B.}~\bibnamefont
  {Sutherland}},\ }\href
  {http://www.sciencedirect.com/science/article/pii/037843638190838X}
  {\bibfield  {journal} {\bibinfo  {journal} {Physica B+C}\ }\textbf {\bibinfo
  {volume} {108}},\ \bibinfo {pages} {1069} (\bibinfo {year}
  {1981})}\BibitemShut {NoStop}%
\bibitem [{\citenamefont {Miyahara}\ and\ \citenamefont
  {Ueda}(1999)}]{Miyahara99}%
  \BibitemOpen
  \bibfield  {author} {\bibinfo {author} {\bibfnamefont {S.}~\bibnamefont
  {Miyahara}}\ and\ \bibinfo {author} {\bibfnamefont {K.}~\bibnamefont
  {Ueda}},\ }\href {https://doi.org/10.1103/PhysRevLett.82.3701} {\bibfield
  {journal} {\bibinfo  {journal} {Phys. Rev. Lett.}\ }\textbf {\bibinfo
  {volume} {82}},\ \bibinfo {pages} {3701} (\bibinfo {year}
  {1999})}\BibitemShut {NoStop}%
\bibitem [{\citenamefont {Momoi}\ and\ \citenamefont
  {Totsuka}(2000)}]{Momoi00}%
  \BibitemOpen
  \bibfield  {author} {\bibinfo {author} {\bibfnamefont {T.}~\bibnamefont
  {Momoi}}\ and\ \bibinfo {author} {\bibfnamefont {K.}~\bibnamefont
  {Totsuka}},\ }\href {https://doi.org/10.1103/PhysRevB.62.15067} {\bibfield
  {journal} {\bibinfo  {journal} {Phys. Rev. B}\ }\textbf {\bibinfo {volume}
  {62}},\ \bibinfo {pages} {15067} (\bibinfo {year} {2000})}\BibitemShut
  {NoStop}%
\bibitem [{\citenamefont {Schmidt}\ \emph {et~al.}(2008)\citenamefont
  {Schmidt}, \citenamefont {Dorier}, \citenamefont {L\"auchli},\ and\
  \citenamefont {Mila}}]{Schmidt2008}%
  \BibitemOpen
  \bibfield  {author} {\bibinfo {author} {\bibfnamefont {K.~P.}\ \bibnamefont
  {Schmidt}}, \bibinfo {author} {\bibfnamefont {J.}~\bibnamefont {Dorier}},
  \bibinfo {author} {\bibfnamefont {A.~M.}\ \bibnamefont {L\"auchli}},\ and\
  \bibinfo {author} {\bibfnamefont {F.}~\bibnamefont {Mila}},\ }\href
  {https://doi.org/10.1103/PhysRevLett.100.090401} {\bibfield  {journal}
  {\bibinfo  {journal} {Phys. Rev. Lett.}\ }\textbf {\bibinfo {volume} {100}},\
  \bibinfo {pages} {090401} (\bibinfo {year} {2008})}\BibitemShut {NoStop}%
\bibitem [{\citenamefont {Radtke}\ \emph {et~al.}(2015)\citenamefont {Radtke},
  \citenamefont {Sa{\'u}l}, \citenamefont {Dabkowska}, \citenamefont
  {Salamon},\ and\ \citenamefont {Jaime}}]{Radtke2015}%
  \BibitemOpen
  \bibfield  {author} {\bibinfo {author} {\bibfnamefont {G.}~\bibnamefont
  {Radtke}}, \bibinfo {author} {\bibfnamefont {A.}~\bibnamefont {Sa{\'u}l}},
  \bibinfo {author} {\bibfnamefont {H.~A.}\ \bibnamefont {Dabkowska}}, \bibinfo
  {author} {\bibfnamefont {M.~B.}\ \bibnamefont {Salamon}},\ and\ \bibinfo
  {author} {\bibfnamefont {M.}~\bibnamefont {Jaime}},\ }\href@noop {}
  {\bibfield  {journal} {\bibinfo  {journal} {Proceedings of the National
  Academy of Sciences}\ }\textbf {\bibinfo {volume} {112}},\ \bibinfo {pages}
  {1971} (\bibinfo {year} {2015})}\BibitemShut {NoStop}%
\bibitem [{\citenamefont {Koga}\ and\ \citenamefont
  {Kawakami}(2000)}]{Koga2000}%
  \BibitemOpen
  \bibfield  {author} {\bibinfo {author} {\bibfnamefont {A.}~\bibnamefont
  {Koga}}\ and\ \bibinfo {author} {\bibfnamefont {N.}~\bibnamefont
  {Kawakami}},\ }\href {https://doi.org/10.1103/PhysRevLett.84.4461} {\bibfield
   {journal} {\bibinfo  {journal} {Phys. Rev. Lett.}\ }\textbf {\bibinfo
  {volume} {84}},\ \bibinfo {pages} {4461} (\bibinfo {year}
  {2000})}\BibitemShut {NoStop}%
\bibitem [{\citenamefont {Takushima}\ \emph {et~al.}(2001)\citenamefont
  {Takushima}, \citenamefont {Koga},\ and\ \citenamefont
  {Kawakami}}]{Takushima01}%
  \BibitemOpen
  \bibfield  {author} {\bibinfo {author} {\bibfnamefont {Y.}~\bibnamefont
  {Takushima}}, \bibinfo {author} {\bibfnamefont {A.}~\bibnamefont {Koga}},\
  and\ \bibinfo {author} {\bibfnamefont {N.}~\bibnamefont {Kawakami}},\ }\href
  {https://doi.org/10.1143/JPSJ.70.1369} {\bibfield  {journal} {\bibinfo
  {journal} {J. Phys. Soc. Jpn.}\ }\textbf {\bibinfo {volume} {70}},\ \bibinfo
  {pages} {1369} (\bibinfo {year} {2001})}\BibitemShut {NoStop}%
\bibitem [{\citenamefont {Chung}\ \emph {et~al.}(2001)\citenamefont {Chung},
  \citenamefont {Marston},\ and\ \citenamefont {Sachdev}}]{Chung01}%
  \BibitemOpen
  \bibfield  {author} {\bibinfo {author} {\bibfnamefont {C.~H.}\ \bibnamefont
  {Chung}}, \bibinfo {author} {\bibfnamefont {J.~B.}\ \bibnamefont {Marston}},\
  and\ \bibinfo {author} {\bibfnamefont {S.}~\bibnamefont {Sachdev}},\ }\href
  {https://doi.org/10.1103/PhysRevB.64.134407} {\bibfield  {journal} {\bibinfo
  {journal} {Phys. Rev. B}\ }\textbf {\bibinfo {volume} {64}},\ \bibinfo
  {pages} {134407} (\bibinfo {year} {2001})}\BibitemShut {NoStop}%
\bibitem [{\citenamefont {L{\"a}uchli}\ \emph {et~al.}(2002)\citenamefont
  {L{\"a}uchli}, \citenamefont {Wessel},\ and\ \citenamefont
  {Sigrist}}]{Laeuchli2002}%
  \BibitemOpen
  \bibfield  {author} {\bibinfo {author} {\bibfnamefont {A.}~\bibnamefont
  {L{\"a}uchli}}, \bibinfo {author} {\bibfnamefont {S.}~\bibnamefont
  {Wessel}},\ and\ \bibinfo {author} {\bibfnamefont {M.}~\bibnamefont
  {Sigrist}},\ }\href {https://doi.org/10.1103/PhysRevB.66.014401} {\bibfield
  {journal} {\bibinfo  {journal} {Phys. Rev. B}\ }\textbf {\bibinfo {volume}
  {66}},\ \bibinfo {pages} {014401} (\bibinfo {year} {2002})}\BibitemShut
  {NoStop}%
\bibitem [{\citenamefont {Corboz}\ and\ \citenamefont
  {Mila}(2013)}]{Corboz13_shastry}%
  \BibitemOpen
  \bibfield  {author} {\bibinfo {author} {\bibfnamefont {P.}~\bibnamefont
  {Corboz}}\ and\ \bibinfo {author} {\bibfnamefont {F.}~\bibnamefont {Mila}},\
  }\href {https://doi.org/10.1103/PhysRevB.87.115144} {\bibfield  {journal}
  {\bibinfo  {journal} {Phys. Rev. B}\ }\textbf {\bibinfo {volume} {87}},\
  \bibinfo {pages} {115144} (\bibinfo {year} {2013})}\BibitemShut {NoStop}%
\bibitem [{\citenamefont {Senthil}\ \emph {et~al.}(2004)\citenamefont
  {Senthil}, \citenamefont {Vishwanath}, \citenamefont {Balents}, \citenamefont
  {Sachdev},\ and\ \citenamefont {Fisher}}]{Senthil2004}%
  \BibitemOpen
  \bibfield  {author} {\bibinfo {author} {\bibfnamefont {T.}~\bibnamefont
  {Senthil}}, \bibinfo {author} {\bibfnamefont {A.}~\bibnamefont {Vishwanath}},
  \bibinfo {author} {\bibfnamefont {L.}~\bibnamefont {Balents}}, \bibinfo
  {author} {\bibfnamefont {S.}~\bibnamefont {Sachdev}},\ and\ \bibinfo {author}
  {\bibfnamefont {M.~P.~A.}\ \bibnamefont {Fisher}},\ }\href
  {http://science.sciencemag.org/content/303/5663/1490.abstract} {\bibfield
  {journal} {\bibinfo  {journal} {Science}\ }\textbf {\bibinfo {volume}
  {303}},\ \bibinfo {pages} {1490} (\bibinfo {year} {2004})}\BibitemShut
  {NoStop}%
\bibitem [{\citenamefont {Lee}\ \emph {et~al.}(2019)\citenamefont {Lee},
  \citenamefont {You}, \citenamefont {Sachdev},\ and\ \citenamefont
  {Vishwanath}}]{Lee2019a}%
  \BibitemOpen
  \bibfield  {author} {\bibinfo {author} {\bibfnamefont {J.~Y.}\ \bibnamefont
  {Lee}}, \bibinfo {author} {\bibfnamefont {Y.-Z.}\ \bibnamefont {You}},
  \bibinfo {author} {\bibfnamefont {S.}~\bibnamefont {Sachdev}},\ and\ \bibinfo
  {author} {\bibfnamefont {A.}~\bibnamefont {Vishwanath}},\ }\href
  {https://doi.org/10.1103/PhysRevX.9.041037} {\bibfield  {journal} {\bibinfo
  {journal} {Phys. Rev. X}\ }\textbf {\bibinfo {volume} {9}},\ \bibinfo {pages}
  {041037} (\bibinfo {year} {2019})}\BibitemShut {NoStop}%
\bibitem [{\citenamefont {Yang}\ \emph {et~al.}(2021)\citenamefont {Yang},
  \citenamefont {Sandvik},\ and\ \citenamefont {Wang}}]{Yang2021}%
  \BibitemOpen
  \bibfield  {author} {\bibinfo {author} {\bibfnamefont {J.}~\bibnamefont
  {Yang}}, \bibinfo {author} {\bibfnamefont {A.~W.}\ \bibnamefont {Sandvik}},\
  and\ \bibinfo {author} {\bibfnamefont {L.}~\bibnamefont {Wang}},\ }\href
  {http://arXiv.org/abs/} {\  (\bibinfo {year} {2021})}\BibitemShut {NoStop}%
\bibitem [{\citenamefont {Shi}\ \emph {et~al.}(2019)\citenamefont {Shi},
  \citenamefont {Steinhardt}, \citenamefont {Graf}, \citenamefont {Corboz},
  \citenamefont {Weickert}, \citenamefont {Harrison}, \citenamefont {Jaime},
  \citenamefont {Marjerrison}, \citenamefont {Dabkowska}, \citenamefont
  {Mila},\ and\ \citenamefont {Haravifard}}]{Shi2019}%
  \BibitemOpen
  \bibfield  {author} {\bibinfo {author} {\bibfnamefont {Z.}~\bibnamefont
  {Shi}}, \bibinfo {author} {\bibfnamefont {W.}~\bibnamefont {Steinhardt}},
  \bibinfo {author} {\bibfnamefont {D.}~\bibnamefont {Graf}}, \bibinfo {author}
  {\bibfnamefont {P.}~\bibnamefont {Corboz}}, \bibinfo {author} {\bibfnamefont
  {F.}~\bibnamefont {Weickert}}, \bibinfo {author} {\bibfnamefont
  {N.}~\bibnamefont {Harrison}}, \bibinfo {author} {\bibfnamefont
  {M.}~\bibnamefont {Jaime}}, \bibinfo {author} {\bibfnamefont
  {C.}~\bibnamefont {Marjerrison}}, \bibinfo {author} {\bibfnamefont {H.~A.}\
  \bibnamefont {Dabkowska}}, \bibinfo {author} {\bibfnamefont {F.}~\bibnamefont
  {Mila}},\ and\ \bibinfo {author} {\bibfnamefont {S.}~\bibnamefont
  {Haravifard}},\ }\href {https://doi.org/10.1038/s41467-019-10410-x}
  {\bibfield  {journal} {\bibinfo  {journal} {Nature Communications}\ }\textbf
  {\bibinfo {volume} {10}},\ \bibinfo {pages} {2439} (\bibinfo {year}
  {2019})}\BibitemShut {NoStop}%
\bibitem [{\citenamefont {Steinhardt}\ \emph {et~al.}(2019)\citenamefont
  {Steinhardt}, \citenamefont {Shi}, \citenamefont {Samarakoon}, \citenamefont
  {Dissanayake}, \citenamefont {Graf}, \citenamefont {Liu}, \citenamefont
  {Zhu}, \citenamefont {Marjerrison}, \citenamefont {Batista},\ and\
  \citenamefont {Haravifard}}]{Steinhardt2019}%
  \BibitemOpen
  \bibfield  {author} {\bibinfo {author} {\bibfnamefont {W.~M.}\ \bibnamefont
  {Steinhardt}}, \bibinfo {author} {\bibfnamefont {Z.}~\bibnamefont {Shi}},
  \bibinfo {author} {\bibfnamefont {A.}~\bibnamefont {Samarakoon}}, \bibinfo
  {author} {\bibfnamefont {S.}~\bibnamefont {Dissanayake}}, \bibinfo {author}
  {\bibfnamefont {D.}~\bibnamefont {Graf}}, \bibinfo {author} {\bibfnamefont
  {Y.}~\bibnamefont {Liu}}, \bibinfo {author} {\bibfnamefont {W.}~\bibnamefont
  {Zhu}}, \bibinfo {author} {\bibfnamefont {C.}~\bibnamefont {Marjerrison}},
  \bibinfo {author} {\bibfnamefont {C.~D.}\ \bibnamefont {Batista}},\ and\
  \bibinfo {author} {\bibfnamefont {S.}~\bibnamefont {Haravifard}},\ }\href
  {http://arXiv.org/abs/} {\  (\bibinfo {year} {2019})}\BibitemShut {NoStop}%
\bibitem [{SM()}]{SM}%
  \BibitemOpen
  \href@noop {} {}\bibinfo {note} {See Supplemental Material at XXX for
  Supplemental figures S1-S9.}\BibitemShut {Stop}%
\bibitem [{\citenamefont {Haravifard}\ \emph {et~al.}(2006)\citenamefont
  {Haravifard}, \citenamefont {Dunsiger}, \citenamefont {El~Shawish},
  \citenamefont {Gaulin}, \citenamefont {Dabkowska}, \citenamefont {Telling},
  \citenamefont {Perring},\ and\ \citenamefont {Bon\ifmmode~\check{c}\else
  \v{c}\fi{}a}}]{Haravifard2006}%
  \BibitemOpen
  \bibfield  {author} {\bibinfo {author} {\bibfnamefont {S.}~\bibnamefont
  {Haravifard}}, \bibinfo {author} {\bibfnamefont {S.~R.}\ \bibnamefont
  {Dunsiger}}, \bibinfo {author} {\bibfnamefont {S.}~\bibnamefont
  {El~Shawish}}, \bibinfo {author} {\bibfnamefont {B.~D.}\ \bibnamefont
  {Gaulin}}, \bibinfo {author} {\bibfnamefont {H.~A.}\ \bibnamefont
  {Dabkowska}}, \bibinfo {author} {\bibfnamefont {M.~T.~F.}\ \bibnamefont
  {Telling}}, \bibinfo {author} {\bibfnamefont {T.~G.}\ \bibnamefont
  {Perring}},\ and\ \bibinfo {author} {\bibfnamefont {J.}~\bibnamefont
  {Bon\ifmmode~\check{c}\else \v{c}\fi{}a}},\ }\href
  {https://doi.org/10.1103/PhysRevLett.97.247206} {\bibfield  {journal}
  {\bibinfo  {journal} {Phys. Rev. Lett.}\ }\textbf {\bibinfo {volume} {97}},\
  \bibinfo {pages} {247206} (\bibinfo {year} {2006})}\BibitemShut {NoStop}%
\bibitem [{\citenamefont {Wietek}\ \emph {et~al.}(2019)\citenamefont {Wietek},
  \citenamefont {Corboz}, \citenamefont {Wessel}, \citenamefont {Normand},
  \citenamefont {Mila},\ and\ \citenamefont {Honecker}}]{wietek19}%
  \BibitemOpen
  \bibfield  {author} {\bibinfo {author} {\bibfnamefont {A.}~\bibnamefont
  {Wietek}}, \bibinfo {author} {\bibfnamefont {P.}~\bibnamefont {Corboz}},
  \bibinfo {author} {\bibfnamefont {S.}~\bibnamefont {Wessel}}, \bibinfo
  {author} {\bibfnamefont {B.}~\bibnamefont {Normand}}, \bibinfo {author}
  {\bibfnamefont {F.}~\bibnamefont {Mila}},\ and\ \bibinfo {author}
  {\bibfnamefont {A.}~\bibnamefont {Honecker}},\ }\href
  {https://doi.org/10.1103/PhysRevResearch.1.033038} {\bibfield  {journal}
  {\bibinfo  {journal} {Phys. Rev. Research}\ }\textbf {\bibinfo {volume}
  {1}},\ \bibinfo {pages} {033038} (\bibinfo {year} {2019})}\BibitemShut
  {NoStop}%
\bibitem [{\citenamefont {Wang}\ and\ \citenamefont
  {Batista}(2018)}]{Wang2018}%
  \BibitemOpen
  \bibfield  {author} {\bibinfo {author} {\bibfnamefont {Z.}~\bibnamefont
  {Wang}}\ and\ \bibinfo {author} {\bibfnamefont {C.~D.}\ \bibnamefont
  {Batista}},\ }\href {https://doi.org/10.1103/PhysRevLett.120.247201}
  {\bibfield  {journal} {\bibinfo  {journal} {Phys. Rev. Lett.}\ }\textbf
  {\bibinfo {volume} {120}},\ \bibinfo {pages} {247201} (\bibinfo {year}
  {2018})}\BibitemShut {NoStop}%
\bibitem [{\citenamefont {Moliner}\ \emph {et~al.}(2011)\citenamefont
  {Moliner}, \citenamefont {Rousochatzakis},\ and\ \citenamefont
  {Mila}}]{moliner11}%
  \BibitemOpen
  \bibfield  {author} {\bibinfo {author} {\bibfnamefont {M.}~\bibnamefont
  {Moliner}}, \bibinfo {author} {\bibfnamefont {I.}~\bibnamefont
  {Rousochatzakis}},\ and\ \bibinfo {author} {\bibfnamefont {F.}~\bibnamefont
  {Mila}},\ }\href {https://doi.org/10.1103/PhysRevB.83.140414} {\bibfield
  {journal} {\bibinfo  {journal} {Phys. Rev. B}\ }\textbf {\bibinfo {volume}
  {83}},\ \bibinfo {pages} {140414} (\bibinfo {year} {2011})}\BibitemShut
  {NoStop}%
\bibitem [{\citenamefont {Manmana}\ \emph {et~al.}(2011)\citenamefont
  {Manmana}, \citenamefont {Picon}, \citenamefont {Schmidt},\ and\
  \citenamefont {Mila}}]{Manmana_2011}%
  \BibitemOpen
  \bibfield  {author} {\bibinfo {author} {\bibfnamefont {S.~R.}\ \bibnamefont
  {Manmana}}, \bibinfo {author} {\bibfnamefont {J.-D.}\ \bibnamefont {Picon}},
  \bibinfo {author} {\bibfnamefont {K.~P.}\ \bibnamefont {Schmidt}},\ and\
  \bibinfo {author} {\bibfnamefont {F.}~\bibnamefont {Mila}},\ }\href
  {https://doi.org/10.1209/0295-5075/94/67004} {\bibfield  {journal} {\bibinfo
  {journal} {{EPL} (Europhysics Letters)}\ }\textbf {\bibinfo {volume} {94}},\
  \bibinfo {pages} {67004} (\bibinfo {year} {2011})}\BibitemShut {NoStop}%
\bibitem [{\citenamefont {Dabkowska}\ \emph {et~al.}(2007)\citenamefont
  {Dabkowska}, \citenamefont {Dabkowski}, \citenamefont {Luke}, \citenamefont
  {Dunsiger}, \citenamefont {Haravifard}, \citenamefont {Cecchinel},\ and\
  \citenamefont {Gaulin}}]{Dabkowska2007}%
  \BibitemOpen
  \bibfield  {author} {\bibinfo {author} {\bibfnamefont {H.~A.}\ \bibnamefont
  {Dabkowska}}, \bibinfo {author} {\bibfnamefont {A.~B.}\ \bibnamefont
  {Dabkowski}}, \bibinfo {author} {\bibfnamefont {G.~M.}\ \bibnamefont {Luke}},
  \bibinfo {author} {\bibfnamefont {S.~R.}\ \bibnamefont {Dunsiger}}, \bibinfo
  {author} {\bibfnamefont {S.}~\bibnamefont {Haravifard}}, \bibinfo {author}
  {\bibfnamefont {M.}~\bibnamefont {Cecchinel}},\ and\ \bibinfo {author}
  {\bibfnamefont {B.~D.}\ \bibnamefont {Gaulin}},\ }\href
  {http://www.sciencedirect.com/science/article/pii/S0022024807004253}
  {\bibfield  {journal} {\bibinfo  {journal} {Journal of Crystal Growth}\
  }\textbf {\bibinfo {volume} {306}},\ \bibinfo {pages} {123} (\bibinfo {year}
  {2007})}\BibitemShut {NoStop}%
\bibitem [{\citenamefont {Verstraete}\ and\ \citenamefont
  {Cirac}(2004)}]{verstraete2004}%
  \BibitemOpen
  \bibfield  {author} {\bibinfo {author} {\bibfnamefont {F.}~\bibnamefont
  {Verstraete}}\ and\ \bibinfo {author} {\bibfnamefont {J.~I.}\ \bibnamefont
  {Cirac}},\ }\href {http://arXiv.org/abs/} {\  (\bibinfo {year}
  {2004})}\BibitemShut {NoStop}%
\bibitem [{\citenamefont {Jordan}\ \emph {et~al.}(2008)\citenamefont {Jordan},
  \citenamefont {Or\'us}, \citenamefont {Vidal}, \citenamefont {Verstraete},\
  and\ \citenamefont {Cirac}}]{jordan2008}%
  \BibitemOpen
  \bibfield  {author} {\bibinfo {author} {\bibfnamefont {J.}~\bibnamefont
  {Jordan}}, \bibinfo {author} {\bibfnamefont {R.}~\bibnamefont {Or\'us}},
  \bibinfo {author} {\bibfnamefont {G.}~\bibnamefont {Vidal}}, \bibinfo
  {author} {\bibfnamefont {F.}~\bibnamefont {Verstraete}},\ and\ \bibinfo
  {author} {\bibfnamefont {J.~I.}\ \bibnamefont {Cirac}},\ }\href
  {https://doi.org/10.1103/PhysRevLett.101.250602} {\bibfield  {journal}
  {\bibinfo  {journal} {Phys. Rev. Lett.}\ }\textbf {\bibinfo {volume} {101}},\
  \bibinfo {pages} {250602} (\bibinfo {year} {2008})}\BibitemShut {NoStop}%
\bibitem [{\citenamefont {Nishio}\ \emph {et~al.}(2004)\citenamefont {Nishio},
  \citenamefont {Maeshima}, \citenamefont {Gendiar},\ and\ \citenamefont
  {Nishino}}]{nishio2004}%
  \BibitemOpen
  \bibfield  {author} {\bibinfo {author} {\bibfnamefont {Y.}~\bibnamefont
  {Nishio}}, \bibinfo {author} {\bibfnamefont {N.}~\bibnamefont {Maeshima}},
  \bibinfo {author} {\bibfnamefont {A.}~\bibnamefont {Gendiar}},\ and\ \bibinfo
  {author} {\bibfnamefont {T.}~\bibnamefont {Nishino}},\ }\href
  {http://arXiv.org/abs/} {\  (\bibinfo {year} {2004})}\BibitemShut {NoStop}%
\bibitem [{\citenamefont {Jiang}\ \emph {et~al.}(2008)\citenamefont {Jiang},
  \citenamefont {Weng},\ and\ \citenamefont {Xiang}}]{jiang2008}%
  \BibitemOpen
  \bibfield  {author} {\bibinfo {author} {\bibfnamefont {H.~C.}\ \bibnamefont
  {Jiang}}, \bibinfo {author} {\bibfnamefont {Z.~Y.}\ \bibnamefont {Weng}},\
  and\ \bibinfo {author} {\bibfnamefont {T.}~\bibnamefont {Xiang}},\ }\href
  {https://doi.org/10.1103/PhysRevLett.101.090603} {\bibfield  {journal}
  {\bibinfo  {journal} {Phys. Rev. Lett.}\ }\textbf {\bibinfo {volume} {101}},\
  \bibinfo {pages} {090603} (\bibinfo {year} {2008})}\BibitemShut {NoStop}%
\bibitem [{\citenamefont {Corboz}\ \emph {et~al.}(2010)\citenamefont {Corboz},
  \citenamefont {Or\'us}, \citenamefont {Bauer},\ and\ \citenamefont
  {Vidal}}]{corboz2010}%
  \BibitemOpen
  \bibfield  {author} {\bibinfo {author} {\bibfnamefont {P.}~\bibnamefont
  {Corboz}}, \bibinfo {author} {\bibfnamefont {R.}~\bibnamefont {Or\'us}},
  \bibinfo {author} {\bibfnamefont {B.}~\bibnamefont {Bauer}},\ and\ \bibinfo
  {author} {\bibfnamefont {G.}~\bibnamefont {Vidal}},\ }\href
  {https://doi.org/10.1103/PhysRevB.81.165104} {\bibfield  {journal} {\bibinfo
  {journal} {Phys. Rev. B}\ }\textbf {\bibinfo {volume} {81}},\ \bibinfo
  {pages} {165104} (\bibinfo {year} {2010})}\BibitemShut {NoStop}%
\bibitem [{\citenamefont {Phien}\ \emph {et~al.}(2015)\citenamefont {Phien},
  \citenamefont {Bengua}, \citenamefont {Tuan}, \citenamefont {Corboz},\ and\
  \citenamefont {Or\'us}}]{phien15}%
  \BibitemOpen
  \bibfield  {author} {\bibinfo {author} {\bibfnamefont {H.~N.}\ \bibnamefont
  {Phien}}, \bibinfo {author} {\bibfnamefont {J.~A.}\ \bibnamefont {Bengua}},
  \bibinfo {author} {\bibfnamefont {H.~D.}\ \bibnamefont {Tuan}}, \bibinfo
  {author} {\bibfnamefont {P.}~\bibnamefont {Corboz}},\ and\ \bibinfo {author}
  {\bibfnamefont {R.}~\bibnamefont {Or\'us}},\ }\href
  {https://doi.org/10.1103/PhysRevB.92.035142} {\bibfield  {journal} {\bibinfo
  {journal} {Phys. Rev. B}\ }\textbf {\bibinfo {volume} {92}},\ \bibinfo
  {pages} {035142} (\bibinfo {year} {2015})}\BibitemShut {NoStop}%
\bibitem [{\citenamefont {Corboz}(2016)}]{corboz16b}%
  \BibitemOpen
  \bibfield  {author} {\bibinfo {author} {\bibfnamefont {P.}~\bibnamefont
  {Corboz}},\ }\href {https://doi.org/10.1103/PhysRevB.94.035133} {\bibfield
  {journal} {\bibinfo  {journal} {Phys. Rev. B}\ }\textbf {\bibinfo {volume}
  {94}},\ \bibinfo {pages} {035133} (\bibinfo {year} {2016})}\BibitemShut
  {NoStop}%
\bibitem [{\citenamefont {Corboz}\ \emph {et~al.}(2011)\citenamefont {Corboz},
  \citenamefont {White}, \citenamefont {Vidal},\ and\ \citenamefont
  {Troyer}}]{corboz2011}%
  \BibitemOpen
  \bibfield  {author} {\bibinfo {author} {\bibfnamefont {P.}~\bibnamefont
  {Corboz}}, \bibinfo {author} {\bibfnamefont {S.~R.}\ \bibnamefont {White}},
  \bibinfo {author} {\bibfnamefont {G.}~\bibnamefont {Vidal}},\ and\ \bibinfo
  {author} {\bibfnamefont {M.}~\bibnamefont {Troyer}},\ }\href
  {https://doi.org/10.1103/PhysRevB.84.041108} {\bibfield  {journal} {\bibinfo
  {journal} {Phys. Rev. B}\ }\textbf {\bibinfo {volume} {84}},\ \bibinfo
  {pages} {041108} (\bibinfo {year} {2011})}\BibitemShut {NoStop}%
\bibitem [{\citenamefont {Corboz}\ \emph {et~al.}(2014)\citenamefont {Corboz},
  \citenamefont {Rice},\ and\ \citenamefont {Troyer}}]{corboz14_tJ}%
  \BibitemOpen
  \bibfield  {author} {\bibinfo {author} {\bibfnamefont {P.}~\bibnamefont
  {Corboz}}, \bibinfo {author} {\bibfnamefont {T.~M.}\ \bibnamefont {Rice}},\
  and\ \bibinfo {author} {\bibfnamefont {M.}~\bibnamefont {Troyer}},\ }\href
  {https://doi.org/10.1103/PhysRevLett.113.046402} {\bibfield  {journal}
  {\bibinfo  {journal} {Phys. Rev. Lett.}\ }\textbf {\bibinfo {volume} {113}},\
  \bibinfo {pages} {046402} (\bibinfo {year} {2014})}\BibitemShut {NoStop}%
\bibitem [{\citenamefont {Nishino}\ and\ \citenamefont
  {Okunishi}(1996)}]{nishino1996}%
  \BibitemOpen
  \bibfield  {author} {\bibinfo {author} {\bibfnamefont {T.}~\bibnamefont
  {Nishino}}\ and\ \bibinfo {author} {\bibfnamefont {K.}~\bibnamefont
  {Okunishi}},\ }\href {https://doi.org/10.1143/jpsj.65.891} {\bibfield
  {journal} {\bibinfo  {journal} {J. Phys. Soc. Jpn.}\ }\textbf {\bibinfo
  {volume} {65}},\ \bibinfo {pages} {891} (\bibinfo {year} {1996})}\BibitemShut
  {NoStop}%
\bibitem [{\citenamefont {Or\'us}\ and\ \citenamefont
  {Vidal}(2009)}]{orus2009-1}%
  \BibitemOpen
  \bibfield  {author} {\bibinfo {author} {\bibfnamefont {R.}~\bibnamefont
  {Or\'us}}\ and\ \bibinfo {author} {\bibfnamefont {G.}~\bibnamefont {Vidal}},\
  }\href {https://doi.org/10.1103/PhysRevB.80.094403} {\bibfield  {journal}
  {\bibinfo  {journal} {Phys. Rev. B}\ }\textbf {\bibinfo {volume} {80}},\
  \bibinfo {pages} {094403} (\bibinfo {year} {2009})}\BibitemShut {NoStop}%
\end{thebibliography}%

\end{document}